\documentclass[12pt,a4paper]{article}

\usepackage{amsmath}
\usepackage{amssymb}
\usepackage{epsfig}
\usepackage{graphicx}
\usepackage{cite}

\newcommand{\capdef}{}
\newcommand{\mycaption}[2][\capdef]{\renewcommand{\capdef}{#2}%
       \caption[#1]{{\footnotesize #2}}}
\makeatletter
\renewcommand{\fnum@table}{\textbf{\tablename~\thetable}}
\renewcommand{\fnum@figure}{\textbf{\figurename~\thefigure}}
\makeatother

\newcounter{myenumi}

\renewcommand{\themyenumi}{\roman{myenumi}}
{\end{list}}

\newlength{\myem}
\newcounter{mysubequation}[equation]

\makeatletter
\renewcommand{\section}{\@startsection{section}{1}{0em}{-\baselineskip}%
{\baselineskip}{\normalfont\large\bfseries}}
\renewcommand{\subsection}%
{\@startsection{subsection}{2}{0em}{-0.7\baselineskip}%
{0.7\baselineskip}{\normalfont\bfseries}}
\makeatother

\newcommand{\THK}{\mbox{\sf T2K-II}}
\newcommand{\HK}{\mbox{\sf HK}}

\newcommand{\otht}{$(O^\mathrm{tr},H^\mathrm{tr})$}
\newcommand{\owht}{$(O^\mathrm{wr},H^\mathrm{tr})$}
\newcommand{\othw}{$(O^\mathrm{tr},H^\mathrm{wr})$}
\newcommand{\owhw}{$(O^\mathrm{wr},H^\mathrm{wr})$}

\newcommand{\ie}{{\it i.e.}}

\newcommand{\eg}{{\it e.g.}}

\newcommand{\eq}{Eq.}
\newcommand{\eqs}{Eqs.}

\newcommand{\fig}{Figure}
\newcommand{\Fig}{Figure}

\newcommand{\Ref}{Ref.}
\newcommand{\Refs}{Refs.}
\newcommand{\Sec}{Section}
\newcommand{\Secs}{Sections}

\newcommand{\Tab}{Table}

\newcommand{\stheta}{\sin^22\theta_{13}}
\newcommand{\dma}{\Delta m_{31}^2}
\newcommand{\dms}{\Delta m_{21}^2}
\newcommand{\deltacp}{\delta_\mathrm{CP}}

\newcommand{\bi}{\begin{itemize}}
\newcommand{\ei}{\end{itemize}}

\begin{document}


\renewcommand{\thefootnote}{\alph{footnote}}

\begin{flushright}
MADPH-05-1412\\
YITP-SB-04-69\\
SISSA 3/2005/EP\\
\end{flushright}

\vspace*{1cm}

\renewcommand{\thefootnote}{\fnsymbol{footnote}}
\setcounter{footnote}{-1}

{\begin{center}
{\Large\textbf{Resolving parameter degeneracies in long-baseline
experiments by atmospheric neutrino data}}
\end{center}}
\renewcommand{\thefootnote}{\it\alph{footnote}}

\vspace*{.8cm}

{\begin{center} {{\bf
                Patrick Huber\footnote[1]{\makebox[1.cm]{Email:}
                \sf phuber@physics.wisc.edu},
                Michele Maltoni\footnote[2]{\makebox[1.cm]{Email:}
                \sf maltoni@insti.physics.sunysb.edu} and
                Thomas Schwetz\footnote[3]{\makebox[1.cm]{Email:}
                \sf schwetz@sissa.it}
                }}
\end{center}}
{\it
\begin{center}
       $^{a}$Department of Physics, University of Wisconsin\\
       1150 University Avenue, Madison, WI 53706, USA\\[4mm]
       $^{b}$C.N.~Yang Institute for Theoretical Physics, 
       SUNY at Stony Brook\\
       Stony Brook, NY 11794-3840, USA\\[4mm]
       $^c$Scuola Internazionale Superiore di Studi Avanzati\\
       Via Beirut 2--4, I--34014 Trieste, Italy
\end{center}}

\vspace*{0.5cm}

\begin{abstract}
   In this work we show that the physics reach of a long-baseline (LBL)
   neutrino oscillation experiment based on a superbeam and a megaton water
   Cherenkov detector can be significantly increased if the LBL data are
   combined with data from atmospheric neutrinos (ATM) provided by the same
   detector. ATM data are sensitive to the octant of $\theta_{23}$ and to the
   type of the neutrino mass hierarchy, mainly through three-flavor effects in
   $e$-like events. This allows to resolve the so-called $\theta_{23}$- and
   sign($\dma$)-parameter degeneracies in LBL data. As a consequence it
   becomes possible to distinguish the normal from the inverted neutrino mass
   ordering at $2\sigma$~CL from a combined LBL+ATM analysis if $\stheta
   \gtrsim 0.02$. The potential to identify the true values of $\stheta$ and
   the CP-phase $\deltacp$ is significantly increased through the lifting of
   the degeneracies. These claims are supported by a detailed simulation of
   the T2K (phase~II) LBL experiment combined with a full three-flavor
   analysis of ATM data in the HyperKamiokande detector.
\end{abstract}


\newpage

\renewcommand{\thefootnote}{\arabic{footnote}}
\setcounter{footnote}{0}


\section{Introduction}

Thanks to the out-standing developments in recent neutrino physics a rather
clear picture of neutrino oscillation parameters is emerging. We know that
there are two large angles in the lepton mixing matrix, $\theta_{12}$ and
$\theta_{23}$, and one small angle $\theta_{13}$, and the mass-squared
differences $\dms$ and $|\dma|$ are well determined by the global data from
solar~\cite{solar} atmospheric~\cite{atmospheric}, reactor~\cite{reactor}, and
accelerator~\cite{Aliu:2004sq} neutrino experiments. A recent global analysis
of world neutrino oscillation data can be found \eg\ in
\Ref~\cite{Maltoni:2004ei}.
Despite these enormous achievements there are still important open questions
related to neutrino physics, which have to be addressed by future neutrino
oscillation experiments: 
\begin{enumerate}
\item
What is the value of the mixing angle $\theta_{13}$?
\item
What is the value of the complex phase
$\deltacp$ in the mixing matrix?
\item
Is the neutrino mass ordering
normal ($m_1<m_2 < m_3$) or inverted ($m_3 < m_1 < m_2$), \ie, what is 
the sign of $\dma$? 
\end{enumerate}
In contrast to the ``discovery phase'', which was dominated by natural
neutrino sources such as the sun or the atmosphere, the subsequent generation
of oscillation experiments will be mainly based on man-made neutrinos, where
the neutrino source is well under control. Among the future projects are
conventional beam experiments~\cite{conventional}, new reactor
experiments~\cite{newreactor}, superbeam
experiments~\cite{superbeams,Itow:2001ee}, and eventually experiments based on
a neutrino factory or a beta-beam~\cite{Albright:2004iw}.

A characteristic feature in the analysis of future long-baseline (LBL)
experiments is the presence of so-called {\it parameter degeneracies}, see
\eg,
\Refs~\cite{Barger:2001yr,Freund:2001ui,Huber:2002mx,Donini:2003vz,Aoki:2003kc,Yasuda:2004gu,Minakata:2002qi}. Due
to the inherent three-flavor structure of the oscillation probabilities, for
a given experiment in general several disconnected regions in the
multi-dimensional space of oscillation parameters will be
present. Traditionally these degeneracies are referred to in the following
way:
\begin{itemize}
\item
The {\it intrinsic} or
($\deltacp,\theta_{13}$)-degeneracy~\cite{Koike:2000jf,Burguet-Castell:2001ez}:
For a measurement based on the $\nu_\mu \to \nu_e$ oscillation probability for
neutrinos and anti-neutrinos two disconnected solutions appear in the
($\deltacp,\theta_{13}$) plane.
\item
The {\it hierarchy} or sign($\dma$)-degeneracy~\cite{Minakata:2001qm}: The
two solutions corresponding to the two signs of $\dma$ appear in general at
different values of $\deltacp$ and $\theta_{13}$.
\item
The {\it octant} or $\theta_{23}$-degeneracy~\cite{Fogli:1996pv}: Since LBL
experiments are sensitive mainly to $\sin^22\theta_{23}$ it is difficult to
distinguish the two octants $\theta_{23} < \pi/4$ and $\theta_{23} > \pi/4$.
Again, the solutions corresponding to $\theta_{23}$ and $\pi/2 - \theta_{23}$
appear in general at different values of $\deltacp$ and $\theta_{13}$.
\end{itemize}
This leads to an eight-fold ambiguity in the determination of $\theta_{13}$
and $\deltacp$~\cite{Barger:2001yr}, and hence degeneracies provide a serious
limitation in the ability to answer the questions 1 and 2 above. Moreover, the
fact that one speaks of a ``sign($\dma$)-degeneracy'' illustrates that
answering question 3 is difficult: determining the neutrino mass ordering is
equivalent to resolving the sign($\dma$)-degeneracy. 
Several methods to resolve these degeneracies have been proposed, among them
the combination of experiments at various baselines and/or
$(L/E)$-values~\cite{Huber:2002rs,Huber:2003ak,Minakata:2001qm,Barger:2001yr,Cervera:2000kp,Barger:2002rr},
the use of spectral information~\cite{Freund:2001ui,Burguet-Castell:2003vv},
the combination of $\nu_e \to \nu_\mu$ and $\nu_e \to \nu_\tau$ oscillation
channels~\cite{silver}, and the combination of LBL and reactor
experiments~\cite{Minakata:2002jv,Minakata:2003wq,Huber:2003pm,Huber:2004ug,McConnel:2004bd}.

In the present work we discuss a new possibility to resolve the LBL parameter
degeneracies, based on the data from atmospheric neutrinos. It is known that
atmospheric neutrinos are in principle sensitive to $\theta_{13}$ and the
neutrino mass hierarchy due to earth matter effects~\cite{msw} in the $e$-like
events~\cite{Petcov:1998su,Akhmedov:1998ui,Akhmedov:1998xq,chizhov,Bernabeu:2001xn,Bernabeu:2003yp,kajita}.
These effects are most pronounced in the multi-GeV energy range and for large
zenith angles, corresponding to neutrino trajectories crossing the earth
mantle or the mantle and the core. In addition effects from the solar
parameters $\theta_{12}$ and $\dms$ on $e$-like events in the sub-GeV energy
range provide sensitivity to the octant of
$\theta_{23}$~\cite{Kim:1998bv,Peres:1999yi,Gonzalez-Garcia:2004cu}, and in
principle even on $\deltacp$~\cite{Peres:2003wd}. For a recent discussion of
sub-leading effects in atmospheric neutrino oscillations see Ref.~\cite{rccn}.
It turns out, however, that atmospheric (ATM) data on its own can never
compete with LBL in many respects, such as the determination of
$\theta_{13},\deltacp,|\dma|$, or $\sin^22\theta_{23}$. One reason is that
they are limited by systematical uncertainties, \eg\ from the neutrino
fluxes. However, as we will show in the following, due to the effects
mentioned above ATM data can break the sign($\dma$)- and
$\theta_{23}$-degeneracies in LBL data, and hence the combined analysis of LBL
and ATM leads to significant synergies.

An important part of the future neutrino program are gigantic water Cherenkov
detectors at the megaton scale. Apart from serving as detector for LBL
experiments such a facility will provide unprecedented opportunities for
proton decay, solar and atmospheric neutrino experiments, as well as for the
detection of supernova and other astrophysical neutrinos. The projects under
discussion are UNO~\cite{uno} in the US, a megaton detector in the Frejus
laboratory~\cite{frejus} in Europe, and the HyperKamiokande
project~\cite{Itow:2001ee,kajita} in Japan. If a LBL experiment with such a
detector will be built atmospheric neutrino data come for free. Therefore, our
method provides a very efficient possibility to resolve the parameter
degeneracies from LBL data, in contrast to the previously discussed methods,
which in general are based on the combination of two or more expensive
experiments.

In the following we will illustrate how the LBL+ATM combination works by
considering the phase~II of the T2K experiment (\THK)~\cite{Itow:2001ee},
assuming a 4~MW superbeam produced at the J-PARC accelerator, and the 1~Mt
HyperKamiokande (\HK) detector serving as the far detector for the LBL
experiment as well as providing the high statistics atmospheric neutrino data.
Let us note that similar results are expected also for the other megaton
detector proposals. Furthermore, if huge magnetized iron detectors are
available one expects similar synergies between LBL and ATM data. For these
type of detectors the ability to distinguish between neutrinos and
anti-neutrinos will increase the potential of ATM data even
further~\cite{magnetized}. In the present work we stick to the \HK\ water
Cherenkov detector, the potential of magnetized detectors will be considered
elsewhere.

The outline of the paper is as follows. In \Sec~\ref{sec:technicalities} we
give some technical details of our simulation of the \THK\ experiment
(\Sec~\ref{sec:t2k}) and of the atmospheric neutrino analysis
(\Sec~\ref{sec:atm}), and we describe the statistical methods used in the
following (\Sec~\ref{sec:statistics}). In \Sec~\ref{sec:degeneracies} we
illustrate the LBL/ATM complementarity by discussing the effects of parameter
degeneracies for the LBL data (\Sec~\ref{sec:qual-lbl}), whereas in
\Secs~\ref{sec:qual-atm-th23} and \ref{sec:qual-atm-hier} we consider the
effects in ATM data which allow to resolve the octant and the sign($\dma$)
degeneracies, respectively. In \Sec~\ref{sec:resolve} we investigate in detail
the ability to exclude degenerate solutions with LBL+ATM data by performing a
systematic scan of the parameter space of $\theta_{13}$, $\theta_{23}$ and
$\deltacp$. In \Sec~\ref{sec:hierarchy} we discuss the potential to identify
the neutrino mass ordering, for the special case $\theta_{23} = \pi/4$
(\Sec~\ref{sec:hier-max}) as well as for the general case
(\Sec~\ref{sec:hier-nonmax}). In \Sec~\ref{sec:octant-th13} we show how the
sensitivity to $\theta_{13}$ is improved by resolving the
$\theta_{23}$-degeneracy by ATM data, and we summarize our results in
\Sec~\ref{sec:conclusions}.


\section{Description of the experiments and statistical analysis}
\label{sec:technicalities}

\subsection{The \THK\ long-baseline experiment}
\label{sec:t2k}

In the following the label ``LBL'' refers to the phase~II of the T2K
experiment (\THK)~\cite{Itow:2001ee}. We are assuming a high luminosity
superbeam with mean neutrino energy of 0.76~GeV, produced with a target power
of 4~MW at the J-PARC accelerator. The neutrinos are detected at a 1~Mt water
Cherenkov detector, HyperKamiokande (\HK), at a baseline of 295~km and an
off-axis angle of $2^\circ$. We consider 2 years running time with a neutrino
beam, and 6 years with anti-neutrinos, such that comparable event numbers are
obtained for neutrinos and anti-neutrinos.

Our simulation of the \THK\ experiment is performed by using the GLoBES
software package~\cite{globes}. We take into account realistic neutrino
fluxes, detection cross sections, energy resolution, and
efficiencies~\cite{Itow:2001ee}. This experimental information is folded with
the three-flavor oscillation probability, fully taking into account the earth
matter effect. We consider all available information, from $\nu_\mu\to \nu_e$
appearance, as well as $\nu_\mu$ disappearance channels. The signal events are
given by $\nu_e$ and $\nu_\mu$ charged current (CC) interactions,
respectively. We divide the signal into the total event rate, where the full
CC cross section is used, and into the energy spectrum with free
normalization, where only events from quasi-elastic scattering are used, since
non-quasi-elastic events do not allow to reconstruct the neutrino energy.
Various backgrounds such as a $\nu_e$ contamination of the beam, $\nu_\mu$
neutral current events, and misidentification of muon neutrinos as electron
neutrinos are taken into account, using information given in
\Ref~\cite{Itow:2001ee}. We list the numbers of signal and background events
expected in \THK\ for typical oscillation parameters in
\Tab~\ref{tab:eventnumbers}. More details of our \THK\ analysis can be found
in \Ref~\cite{Huber:2002mx}.

\begin{table}
\centering
\begin{tabular}{|l|r|r|l|r|r|}
\hline\hline
\multicolumn{3}{|c}{The \THK\ LBL experiment}&
\multicolumn{3}{|c|}{The \HK\ ATM experiment (9 Mt yrs)}\\
\hline
         & $\nu$ (2 Mt yrs) & $\bar\nu$ (6 Mt yrs) & 
         & $\nu$ & $\bar\nu$ \\
\hline
$\nu_\mu\to\nu_e$ signal       &  21\,300 & 16\,000 &
$e$-like sub-GeV &\hspace*{1mm}  239\,000 & 58\,000 \\ 
$\nu_\mu\to\nu_e$ background   &   2\,140 &  3\,260 &
$e$-like multi-GeV             &  52\,700 & 18\,100 \\ 
$\nu_\mu\to\nu_\mu$ signal     &  73\,200 & 75\,600 &
$\mu$-like sub-GeV             & 232\,000 & 66\,200 \\ 
$\nu_\mu\to\nu_\mu$ background &      340 &     320 &
$\mu$-like multi-GeV           & 108\,000 & 49\,100 \\ 
&&& 
upward going $\mu$             & 127\,000 & 65\,400 \\ 
\hline\hline
\end{tabular}
\mycaption{Number of events in the LBL and ATM experiments considered in our
  work for the oscillation parameters $\stheta = 0.05$, $\sin^2\theta_{23} =
  0.5$, $\sin^2\theta_{12} = 0.3$, $\deltacp = 0$, $\dms = 8.1 \times
  10^{-5}$~eV$^2$, and $\dma = 2.2 \times 10^{-3}$~eV$^2$.}
\label{tab:eventnumbers}
\end{table}

\subsection{The \HK\ atmospheric neutrino experiment}
\label{sec:atm}

To simulate atmospheric neutrino data in the \HK\ detector we follow closely
\Ref~\cite{Gonzalez-Garcia:2004cu}. A crucial element of the analysis is to
take into account the full three-flavor oscillation probability, including
earth matter effects, as well as oscillations induced by the ``solar'' mass
splitting $\dms$ (for other three-flavor analyses see
\Refs~\cite{Yasuda:1996hn,Fogli:2001it,Gonzalez-Garcia:2002mu,lisitalk}).
In our analysis we consider charged current data, divided into sub-GeV and
multi-GeV $e$-like and $\mu$-like contained event samples (each grouped into
10~bins in zenith angle), as well as stopping (5~angular bins) and
through-going (10~angular bins) up-going muon events.

Details of our statistical analysis can be found in the Appendix of
Ref.~\cite{Gonzalez-Garcia:2004wg}. Together with the statistical errors, we
consider theoretical as well as systematical uncertainties, where theoretical
uncertainties are uncertainties in the original atmospheric neutrino fluxes
and in the cross-sections.  We are using the atmospheric neutrino fluxes from
Ref.~\cite{Honda:2004yz}, and flux uncertainties include total normalization
errors (20\%) allowing for an independent fluctuation of neutrino and
anti-neutrino fluxes (5\%) as well as $\nu_\mu$ and $\nu_e$ fluxes (5\%),
spectral uncertainty of the fluxes (``tilt'' factor), and an uncertainty on
the zenith angle dependence which induces an error in the up/down asymmetry of
events (5\%).
We also include independent normalization errors for the different
contributions to the interaction cross section: quasi-elastic scattering
(15\%), single pion production (15\%), and deep inelastic scattering (15\% for
contained events and 10\% for upward-going muons).
Moreover, we include as systematical errors experimental uncertainties
associated with the simulation of the hadronic interactions, the particle
identification procedure, the ring-counting procedure, the fiducial volume
determination, the energy calibration, the relative normalization between
partially-contained and fully-contained events, the track reconstruction of
up-going muons, the detection efficiency of up-going muons, and the
stopping/through-going separation.\footnote{The impact of these theoretical
and systematical uncertainties on the performance of future atmospheric
neutrino experiments has been investigated in some detail in
\Ref~\cite{Gonzalez-Garcia:2004cu}. See also \Ref~\cite{rccn}.}

The current atmospheric neutrino data sample from SuperKamiokande (SK-I)
consists of 1489 days of data (contained events) with a detector mass of
22.5~kt, which gives roughly $90\,\mathrm{kt\,yrs}$ of data. In this work we
are considering a data taking period for the LBL experiment of 8~years. We
assume that the 1~Mt \HK\ detector will be finished one year before the \THK\
beam, and hence $9\,\mathrm{Mt\,yrs}$ of atmospheric neutrino data will be
available. For the atmospheric data sample used in our analysis, which we will
denote in the following by the label ``ATM'', we scale the present SK-I sample
(1489~days contained events, 1657~days of stopping, and 1678~days of
through-going muons) by a factor 100. Event numbers for various ATM data
samples for typical oscillation parameters are given in
\Tab~\ref{tab:eventnumbers}. 


\subsection{Details of the statistical analysis}
\label{sec:statistics}

In order to investigate the potential of the experiments described in the
previous sub-sections we adopt the standard method for analyzing future
experiments. First, artificial ``data'' is simulated by calculating event
numbers for LBL and ATM for some ``true values''
$\boldsymbol{\theta}^\mathrm{true}$ for the oscillation parameters
$\boldsymbol{\theta} = (\theta_{13}, \theta_{12}, \theta_{23},\deltacp, \dms,
\dma)$. Then a $\chi^2$-analysis of these ``data'' is performed to extract
allowed regions for the oscillation parameters. It is important to note that
in general the results will depend on the values adopted for
$\boldsymbol{\theta}^\mathrm{true}$. For all calculations we will use for
$\theta_{12}^\mathrm{true},(\dms)^\mathrm{true}$ and $(\dma)^\mathrm{true}$
the best fit values obtained in \Ref~\cite{Maltoni:2004ei},
\begin{equation}
\label{eq:true-values}
\sin^2\theta_{12} = 0.3 \,,\quad
\dms = 8.1 \times 10^{-5}\,\mathrm{eV}^2\,,\quad
|\dma| = 2.2 \times 10^{-3}\,\mathrm{eV}^2\,.
\end{equation}
We do not expect any significant changes of our results if these parameters
are varied within the present allowed ranges~\cite{Maltoni:2004ei}.  However,
we adopt various values for $\theta_{13}^\mathrm{true},
\theta_{23}^\mathrm{true}$, $\deltacp^\mathrm{true}$, and sign($\dma$), and
show our results as a function of these parameters.  Furthermore, since
neither LBL nor ATM data allow an accurate determination of $\theta_{12}$ and
$\dms$ we assume that these two parameters are known with an uncertainty of
10\% from solar and reactor neutrino experiments.

If one is interested in the allowed range for a certain parameter $\xi \in
\boldsymbol{\theta}$ for a given choice of
$\boldsymbol{\theta}^\mathrm{true}$, the function
$\chi^2(\boldsymbol{\theta}^\mathrm{true}; \boldsymbol{\theta})$ has to be
minimized with respect to all other parameters $\boldsymbol{\theta}$ except
$\xi$ to take into account the correlations and degeneracies between
parameters. This minimization is performed by using the GLoBES software
package~\cite{globes}, which has been generalized in order to include the
atmospheric neutrino code. Let us stress that for both data samples, LBL as
well as ATM, a full three-flavor analysis including matter effects is
performed.  The only approximation is to neglect the (very weak) dependence on
$\theta_{12}$. Since LBL depends in leading order only on the product
$\sin2\theta_{12}\dms$ fixing $\theta_{12}$ does not introduce any error, as
long as the dependence on $\dms$ is properly taken into account, and for ATM
varying $\theta_{12}$ in the allowed range is expected to have a very small
impact. Apart from this simplification the full parameter dependence of both
data sets has been taken into account. However, we assume that LBL and ATM
data are statistically independent. This might not be completely correct since
both data sets are based on the same detector, and hence, uncertainties
related to the detection process can introduce correlations between LBL and
ATM data.  Although such effects will have to be included eventually in the
analysis of {\it real} data, we expect that such correlations introduce only
very minor corrections to our present results.


\section{Complementarity between LBL and ATM data}
\label{sec:degeneracies}

\subsection{Parameter degeneracies and the \THK\ experiment}
\label{sec:qual-lbl}

In this section we discuss the problem of parameter degeneracies for the \THK\
experiment. First we note that for this particular experiment the $(\deltacp,
\theta_{13})$-degeneracy does not occur. It is known that for experiments
operating at the oscillation maximum of $\dma$ the second solution in the
$(\deltacp, \theta_{13})$ plane can be disfavored
efficiently~\cite{Barger:2001yr,Huber:2002mx}. 
Moreover, spectral information is important, since $\deltacp$-dependent and
$\deltacp$-independent terms in the oscillation probability show a different
energy dependence (compare \eq~(\ref{eq:emuprob}) later in this section). Thus
it is difficult to leave the spectrum unchanged when $\deltacp$ is varied, and
the $(\deltacp,\theta_{13})$-degenerate solution, which implies a different
value of $\deltacp$, is disfavored. An illustration of the relevance of
spectral information for the $(\deltacp,\theta_{13})$-degeneracy in the \THK\
experiment can be found in \Fig~5 of \Ref~\cite{Huber:2002mx}.

\begin{figure}[t!]
   \centering
   \includegraphics[width=0.95\textwidth]{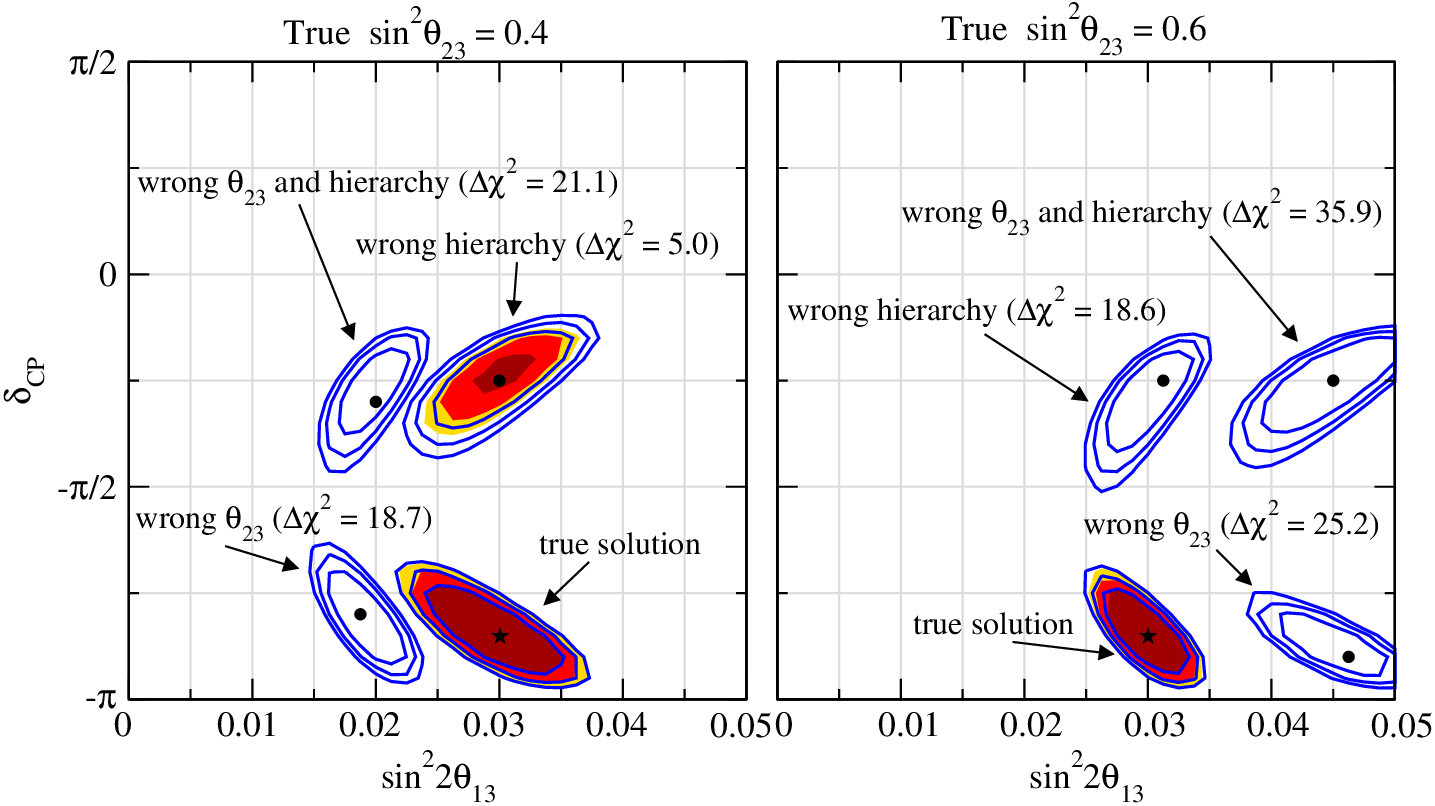}
   \mycaption{Allowed regions in the $(\stheta,\deltacp)$ plane at 2$\sigma$,
   99\%, and 3$\sigma$ CL (2~dof) of the true and all degenerate solutions for
   $\stheta^\mathrm{true} = 0.03$, $\deltacp^\mathrm{true} = -0.85\pi$, and
   $\sin^2\theta_{23}^\mathrm{true} = 0.4$ (left) and
   $\sin^2\theta_{23}^\mathrm{true} = 0.6$ (right). The solid curves
   correspond to LBL data only, and the shaded regions correspond to LBL+ATM
   data. The true best fit point is marked with a star, the best fit points of
   the degenerate solutions are marked with dots, and the corresponding
   $\Delta\chi^2$-values of LBL+ATM data are given in the figure. The true
   mass ordering is the normal hierarchy.}
   \label{fig:d-th-nonmax-2}
\end{figure}

Therefore, in the case of \THK\ we are confronted only with the sign($\dma$)-
and $\theta_{23}$-degeneracies. Apart from the incapacity to determine the
neutrino mass ordering and the octant of $\theta_{23}$ this leads to a
four-fold ambiguity in the determination of $\theta_{13}$ and $\deltacp$.  The
impact of the degeneracies can be appreciated in \Fig~\ref{fig:d-th-nonmax-2},
where the solid curves show the allowed regions from LBL data in the
$(\stheta,\deltacp)$ plane for an example-point with the true values $\stheta
= 0.03$, $\deltacp = -0.85\pi$, and non-maximal values of
$\theta_{23}^\mathrm{true}$. Apart from the true solution, three degenerate
regions are present, corresponding to the wrong octant of $\theta_{23}$, the
wrong sign of $\dma$, and the wrong octant as well as the wrong hierarchy. Let
us introduce the following abbreviations to denote these four solutions:
\begin{equation}\label{eq:solutions}
\begin{array}{l@{\qquad}l}
\text{\otht} & \text{true solution} \\
\text{\othw} & \text{true octant of $\theta_{23}$ and wrong hierarchy}\\
\text{\owht} & \text{wrong octant of $\theta_{23}$ and true hierarchy}\\
\text{\owhw} & \text{wrong octant of $\theta_{23}$ and wrong hierarchy}
\end{array}
\end{equation}

A qualitative understanding of the degenerate solutions can be obtained from
the approximate formula for the $\nu_\mu\to\nu_e$ appearance probability in
vacuum~\cite{Cervera:2000kp}
\begin{eqnarray}
P_{\nu_\mu \rightarrow \nu_e} 
& \simeq & \sin^2 2\theta_{13} \, \sin^2 \theta_{23}
\sin^2 {\Delta} \nonumber \\
& + &  \alpha\; \sin 2\theta_{13} \, \sin 2\theta_{12} \sin 2\theta_{23}
\, \Delta \sin{\Delta} \cos(\Delta \pm \deltacp) \label{eq:emuprob}\\
&+& \alpha^2 \, \cos^2 \theta_{23} \sin^2 2\theta_{12} \, \Delta^2
\nonumber 
\end{eqnarray} 
with $\Delta \equiv \Delta m^2_{31}L / (4E_\nu)$ and $\alpha \equiv
\Delta m^2_{21} / \Delta m^2_{31}$. The sign in the second term is
`$+$' for neutrinos and `$-$' for anti-neutrinos. Since in the \THK\
experiment matter effects are small this expression for the
probability suffices to obtain a qualitative understanding of most of
the effects in LBL data presented throughout this work. 

Following \Ref~\cite{Barger:2001yr} the location of the
$\theta_{23}$-degenerate solution \owht\ can be estimated from
\eq~(\ref{eq:emuprob}) in the following way. Since the \THK\ experiment
operates close to the oscillation maximum it is a good approximation
to use $\Delta \approx \pi/2$. Furthermore, for values of $\stheta
\gtrsim 0.01$ the $\alpha^2$ term can be neglected. Under these
approximation solving the two equations 
\begin{eqnarray}
P_{\nu_\mu\to\nu_e}( \theta_{13},\deltacp, \theta_{23}) &=& 
P_{\nu_\mu\to\nu_e}( \theta_{13}',\deltacp', \pi/2 - \theta_{23})
\nonumber\\
P_{\bar\nu_\mu\to\bar\nu_e}( \theta_{13},\deltacp, \theta_{23}) &=& 
P_{\bar\nu_\mu\to\bar\nu_e}( \theta_{13}',\deltacp', \pi/2 -\theta_{23}) 
\end{eqnarray}
leads to 
\begin{eqnarray}
\stheta' &\approx& \stheta \, \tan^2\theta_{23} 
\label{eq:fake-th13-th23}\\
\sin\deltacp' &\approx& \sin\deltacp \, \tan\theta_{23} 
\label{eq:fake-delta-th23}
\end{eqnarray}
for the parameter values of the wrong-$\theta_{23}$ solution. For the example
shown in \fig~\ref{fig:d-th-nonmax-2} \eq~(\ref{eq:fake-th13-th23}) gives
$\stheta' = 0.02$ for $\sin^2\theta_{23}^\mathrm{true} = 0.4$, and $\stheta' =
0.045$ for $\sin^2\theta_{23}^\mathrm{true} = 0.6$.  Furthermore,
\eq~(\ref{eq:fake-delta-th23}) gives $\deltacp' = -0.81\pi$ for
$\sin^2\theta_{23}^\mathrm{true} = 0.4$, and $\deltacp' = 0.88\pi$ for
$\sin^2\theta_{23}^\mathrm{true} = 0.6$. These numbers are in good agreement
with the actual fit shown in \fig~\ref{fig:d-th-nonmax-2}. One observes that
the $\theta_{23}$-degeneracy has a strong impact on the measurement of
$\stheta$, however, it hardly affects the determination of
$\deltacp$~\cite{Barger:2001yr,Huber:2002mx}; the \owht-solution occurs
practically at the same value of $\deltacp$ as the true one.

This is in contrast to the sign($\dma$)-degeneracy, which in general leads to
a severe ambiguity for $\deltacp$, whereas the $\stheta$ measurement is
essentially unaffected.  Following \Ref~\cite{Minakata:2001qm} we observe from
\eq~(\ref{eq:emuprob}) that only the term in the second line is affected by
changing the sign of $\dma$. This term transforms as
\begin{equation}
\alpha\, \Delta \, \sin\Delta \cos(\Delta \pm \deltacp) \to 
- \alpha\, \Delta \, \sin\Delta \cos(-\Delta \pm \deltacp)
\end{equation}
under $\dma \to -\dma$. Therefore, the full probability stays
invariant under this transformation if $\deltacp$ is adjusted such
that $\cos(-\Delta \pm \deltacp) = -\cos(\Delta \pm \deltacp')$, which
can be achieved for 
\begin{equation}\label{eq:fake-delta-h}
\deltacp' = \pi - \deltacp\,. 
\end{equation}
According to this reasoning we would obtain in the example plotted in
\Fig~\ref{fig:d-th-nonmax-2} the value $\deltacp' = 1.85\pi \,\hat=\,
-0.15\pi$. Although in this case the accuracy is not excellent,
\eq~(\ref{eq:fake-delta-h}) still provides a rough method to estimate the
location of the \othw-solution in $\deltacp$ (for more refined methods see
\eg\ \Refs~\cite{patrickthesis,Donini:2003vz}). 
Finally, the values of $\stheta$ and $\deltacp$ corresponding to the combined
$\theta_{23}$ and sign($\dma$)-degeneracy \owhw\ can be estimated by applying
simultaneously \eqs~(\ref{eq:fake-th13-th23}) and (\ref{eq:fake-delta-h}).

This four-fold degeneracy can be lifted to large extent if LBL data is
combined with data from atmospheric neutrinos. We observe from
\Fig~\ref{fig:d-th-nonmax-2} that the degenerate solutions corresponding to the
wrong octant of $\theta_{23}$ are highly disfavored by the inclusion of ATM
data, at the level of $\Delta \chi^2 \gtrsim 20$. Furthermore, also the
solution with the wrong mass ordering gets disfavored in the combined
analysis, although in this case the ability to resolve the degeneracy is more
subtle. In the following subsections we discuss the relevant features of ATM
data to resolve the degeneracies in LBL data.


\subsection{The sensitivity of ATM data to the octant of $\theta_{23}$}
\label{sec:qual-atm-th23}

The \THK\ experiment will provide a very precise determination of
$\sin^22\theta_{23}$ thanks to the large statistics data from the $\nu_\mu$
disappearance channel (compare \Tab~\ref{tab:eventnumbers}). The relative
accuracy at 2$\sigma$ is expected to be better than 1\%. Despite this
impressive performance on $\sin^22\theta_{23}$ there are some subtleties
related to the measurement of $\sin^2\theta_{23}$ (see, \eg,
\Refs~\cite{Antusch:2004yx,Minakata:2004pg}). Especially, if $\theta_{23}$
deviates from $\pi/4$ it will be impossible to distinguish between the two
solutions at $\theta_{23}$ and $\pi/2 - \theta_{23}$.\footnote{For theoretical
expectations for the deviations of $\theta_{23}$ from the maximal value see
\eg\ \Refs~\cite{Antusch:2004yx,Antusch:2003kp} and references therein.}
Although the determination of $\sin^2 2\theta_{23}$ from atmospheric data is
significantly less precise than from LBL they provide the very interesting
ability to distinguish between the two octants of $\theta_{23}$.

The sensitivity of atmospheric data to the deviation of $\theta_{23}$ from
$\pi/4$ follows mainly from effects of the solar mass splitting $\dms$ for
$e$-like events in the sub-GeV region, see \eg\
\Refs~\cite{Kim:1998bv,Peres:1999yi,Peres:2003wd,Gonzalez-Garcia:2004cu}.  The
excess of $e$-like events can be written as
\begin{equation}\label{eq:excess-sub}
\epsilon_e^\mathrm{sub} \equiv \frac{N_e}{N_e^0} - 1 \approx
\left( r \, \cos^2\theta_{23} - 1 \right) \langle P_{21}^{2\nu}
\rangle \,.
\end{equation}
Here $N_e \, (N_e^0)$ is the number of $e$-like events with (without)
oscillations, $\langle P_{21}^{2\nu}\rangle$ is the averaged two neutrino
oscillation probability given by the solar parameters $\Delta m_{21}^2$ and
$\theta_{12}$ including the (weighted) contributions from neutrinos and
anti-neutrinos, and $r \equiv F_\mu^0 / F_e^0$ is the ratio of the initial
muon and electron neutrino fluxes. Since for sub-GeV energies $r\approx 2$ the
effect is suppressed for $\theta_{23} \approx \pi/4$, however it provides a
sensitive measure for deviations from maximal mixing. Given the present LMA
parameters, for $|\sin^2\theta_{23} - 0.5| \approx 0.1$ one expects
$\epsilon_e^\mathrm{sub}$ values of a few percent (see \eg\ Fig.~8 of
\Ref~\cite{Peres:2003wd}). In contrast to LBL data, which is essentially
sensitive only to $\sin^22\theta_{23}$, this effect depends on
$\cos^2\theta_{23}$, and therefore the discrimination between $\theta_{23} >
\pi/4$ and $\theta_{23} < \pi/4$ becomes possible.

\begin{figure}[t!]
   \centering 
    \includegraphics[width=0.5\textwidth]{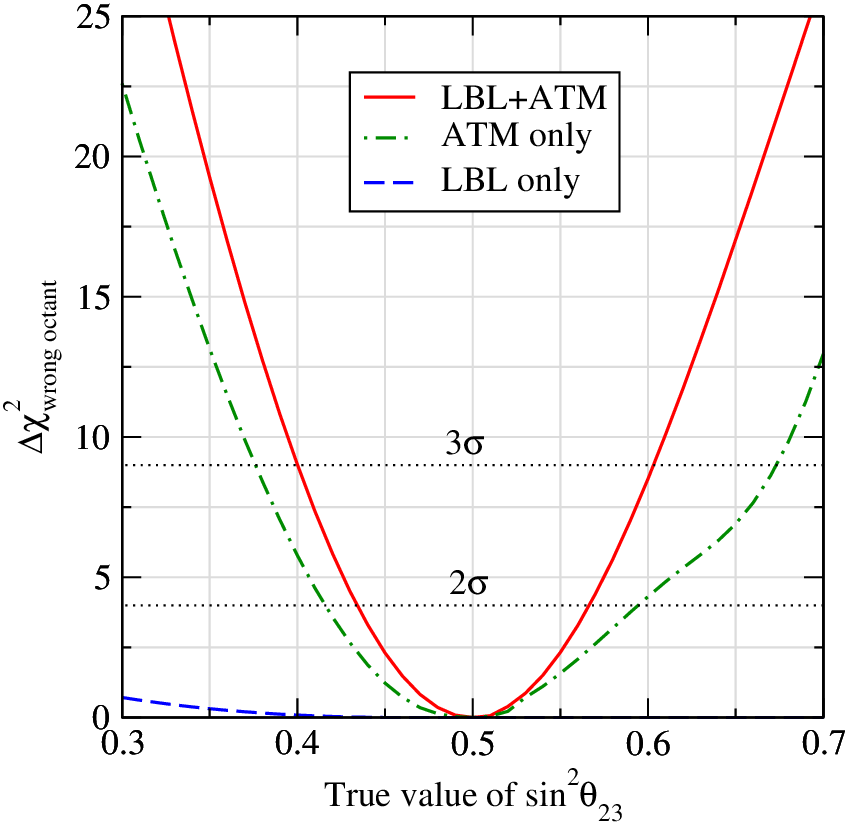}
    \mycaption{$\Delta\chi^2$ of the solution in the wrong octant of
    $\theta_{23}$ as a function of the true value of $\sin^2\theta_{23}$ for
    LBL data only, for ATM data only, and for the LBL+ATM
    combination. Furthermore, we take $\theta_{13}^\mathrm{true} = 0$.}
    \label{fig:wrong-th23}
\end{figure}

Building upon the results of \Ref~\cite{Gonzalez-Garcia:2004cu} we show in
\Fig~\ref{fig:wrong-th23} the difference in $\chi^2$ between the true solution
(with $\chi^2=0$) and the $\chi^2$-minimum in the wrong octant of
$\theta_{23}$ (\owht-solution). It is clear form this figure that LBL data
alone have no sensitivity at all to the octant of $\theta_{23}$, whereas ATM
can reject the fake solution efficiently.  Taking the combination of LBL+ATM
data improves the sensitivity slightly, due to the more precise determination
of other oscillation parameters by the LBL data. Using the combined LBL+ATM
data the wrong octant can be rejected at 3$\sigma$ if $|\sin^2\theta_{23} -
0.5| > 0.1$. Let us note that the results shown in \Fig~\ref{fig:wrong-th23}
do hardly depend on our choice for the true neutrino mass ordering, thanks to
$\theta_{13}^\mathrm{true} = 0$. Non-zero values for
$\theta_{13}^\mathrm{true}$ will be considered in \Sec~\ref{sec:resolve}.


\subsection{The sensitivity of ATM data to the mass hierarchy}
\label{sec:qual-atm-hier}

The determination of the ordering of the neutrino mass states is one of the
most challenging tasks of future neutrino experiments. In long-baseline
experiments the effects of $\theta_{13}$, $\deltacp$ and the sign of $\dma$
are highly entangled, and the determination of sign($\dma$) is probably only
possible through the matter effect induced in an experiment with a very long
baseline plus additional information from other LBL or reactor
experiments~\cite{Barger:2000cp,Lipari:1999wy,Huber:2003pm,Huber:2003ak,Huber:2002rs,Huber:2002mx,Minakata:2001qm,Mena:2004sa}.\footnote{Another
possibility to identify the neutrino mass hierarchy could come from the
observation of neutrinos emitted by a galactic supernova, see \eg\ 
\Ref~\cite{Kachelriess:2004vs} for a recent analysis.}
Therefore, the combined analysis of LBL and ATM data as offered from an
oscillation experiment with a Mt water Cherenkov detector provides a very
interesting possibility to answer the question of the neutrino mass hierarchy.

The sensitivity of atmospheric neutrino data to the neutrino mass hierarchy
comes mainly from the modification of $e$-like multi-GeV events by earth
matter effects for not too small values of
$\stheta$~\cite{Akhmedov:1998xq,Bernabeu:2003yp,Bernabeu:2001xn,chizhov,Petcov:1998su,Akhmedov:1998ui,kajita}.
Similar as in \eq~(\ref{eq:excess-sub}), one finds for the $\theta_{13}$
induced excess of $e$-like multi-GeV events
\begin{equation}\label{eq:excess-multi}
\epsilon_e^\mathrm{multi} \equiv \frac{N_e}{N_e^0} - 1 \approx
\left( r \, \sin^2\theta_{23} - 1 \right) \langle P_{31}^{2\nu}
\rangle \,.
\end{equation}
Now $\langle P_{31}^{2\nu}\rangle$ is an effective two-flavor probability
governed by $\dma$ and $\theta_{13}$, appropriately averaged and including the
(weighted) contributions from neutrinos and anti-neutrinos. The effect is most
pronounced for zenith angles corresponding to neutrino trajectories crossing
the earth mantle, or earth mantle and core, where $\stheta$-effects can be
resonantly enhanced due to matter
effects~\cite{msw,Petcov:1998su,Akhmedov:1998ui}.
In the relevant zenith angle bins $\epsilon_e^\mathrm{multi}$ can reach values
of the order of 10\% (see \eg\ Fig.~5 of \Ref~\cite{Bernabeu:2003yp}).
Qualitatively $\epsilon_e^\mathrm{multi}$ shows the following
behavior~\cite{Bernabeu:2003yp}:
\begin{itemize}
\item
$\epsilon_e^\mathrm{multi}$ vanishes for $\stheta = 0$ and increases
monotonically with $\stheta$.
\item
For the normal hierarchy the resonant matter enhancement occurs for neutrinos,
whereas for the inverted hierarchy it occurs for anti-neutrinos. Since the
event numbers in water Cherenkov detectors are dominated by neutrinos because
of larger cross sections (see \Tab~\ref{tab:eventnumbers}),
$\epsilon_e^\mathrm{multi}$ is larger by a factor of $1.5 - 2$ for the normal
hierarchy than for the inverted one.
\item
Since in the multi-GeV range we have $r\simeq 2.6 - 4.5$, the factor $(r \,
\sin^2\theta_{23} - 1)$ in \eq~(\ref{eq:excess-multi}) suppresses the excess
of $e$-like events for small values $\sin^2\theta_{23} \lesssim 0.4$, whereas
larger values of $\sin^2\theta_{23}$ increase $\epsilon_e^\mathrm{multi}$. In
particular, the effect is enhanced for $\theta_{23} > \pi/4$.
\end{itemize}
Therefore, if the true hierarchy is normal and $\theta_{23} < \pi/4$ there is
only a small excess of $e$-like multi-GeV events, which can be accommodated to
some extent with the inverted hierarchy. For the example point with
$\sin^2\theta_{23}^\mathrm{true} = 0.4$ chosen in the left panel of
\Fig~\ref{fig:d-th-nonmax-2} the wrong hierarchy can be disfavored only with
a $\Delta \chi^2 = 5.0$. In contrast, for $\theta_{23} > \pi/4$ the excess of
$e$-like events is enhanced by the flux-factor $(r \, \sin^2\theta_{23} -
1)$. Therefore, a true normal hierarchy (resonant enhancement for neutrinos)
plus the flux-factor enhancement leads to large values of
$\epsilon_e^\mathrm{multi}$, which cannot be fitted with the inverted
hierarchy (resonant enhancement for the smaller anti-neutrino sample). This
explains the strong rejection of the wrong hierarchy solution in the right
panel of \Fig~\ref{fig:d-th-nonmax-2}, with a $\Delta\chi^2 = 18.6$. We will
discuss the mechanisms relevant for the rejection of the wrong hierarchy in
more detail in \Secs~\ref{sec:resolve} and \ref{sec:hierarchy}.

Finally we note that in addition to $\dms$-effects of
\eq~(\ref{eq:excess-sub}) and $\theta_{13}$-effects of
\eq~(\ref{eq:excess-multi}) also an interference term between the two
contributions is present~\cite{Peres:2003wd} (see also
\Ref~\cite{lisitalk}). It is proportional to $(r \, \sin\theta_{13} \, \sin
2\theta_{23})$ and depends on the CP-phase $\deltacp$. Because of the
different dependence on the flux ratio $r$ the interference term may become
important in cases where the effects governed by \eqs~(\ref{eq:excess-sub})
and (\ref{eq:excess-multi}) are suppressed.


\section{Resolving the degeneracies}
\label{sec:resolve}

\begin{figure}[t!]
   \centering 
   \includegraphics[width=0.58\textwidth]{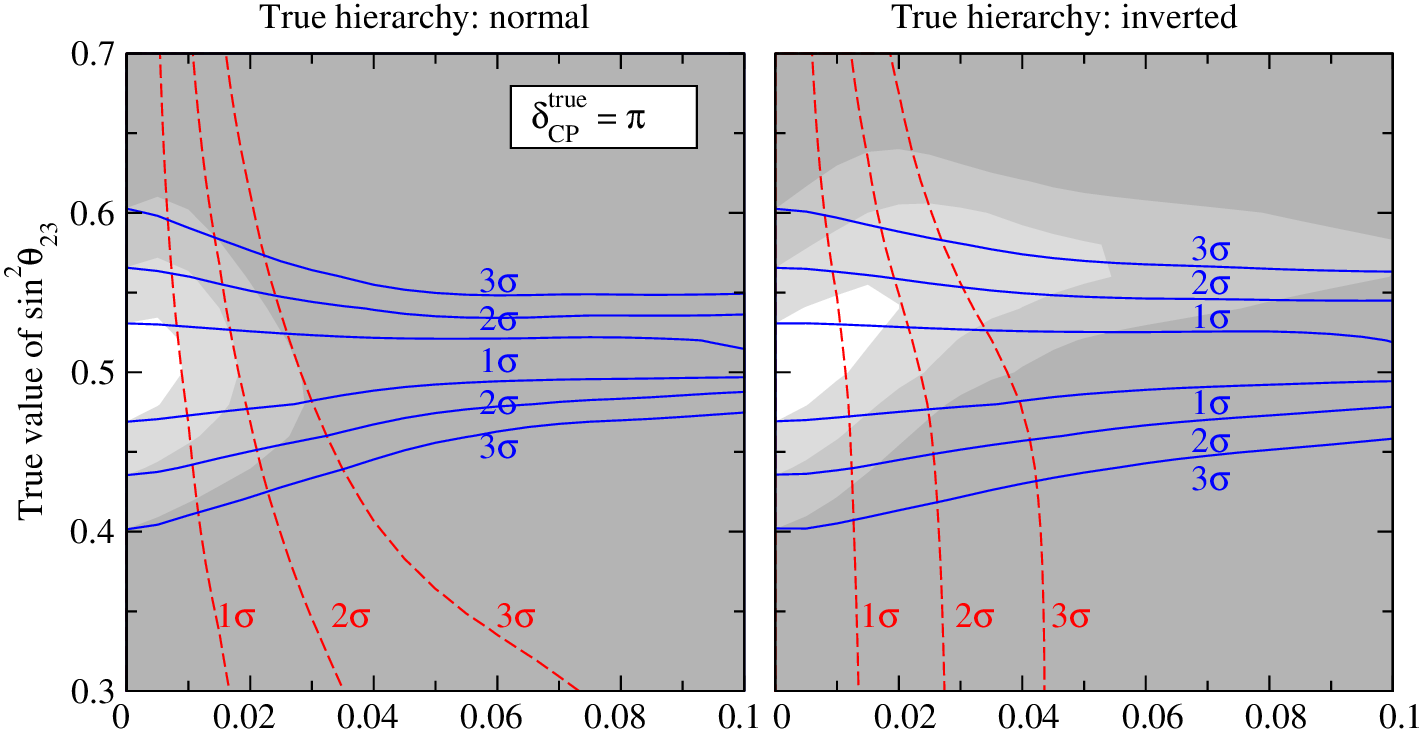}
   \includegraphics[width=0.58\textwidth]{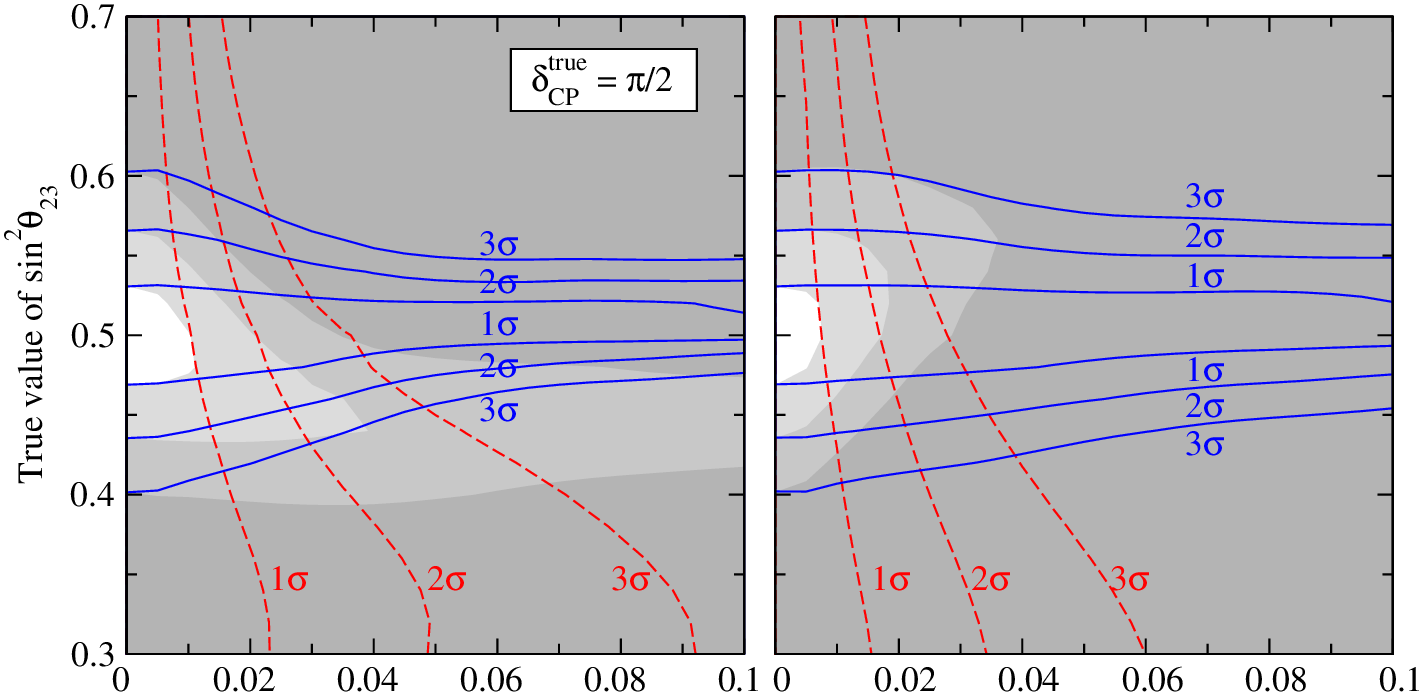}
   \includegraphics[width=0.58\textwidth]{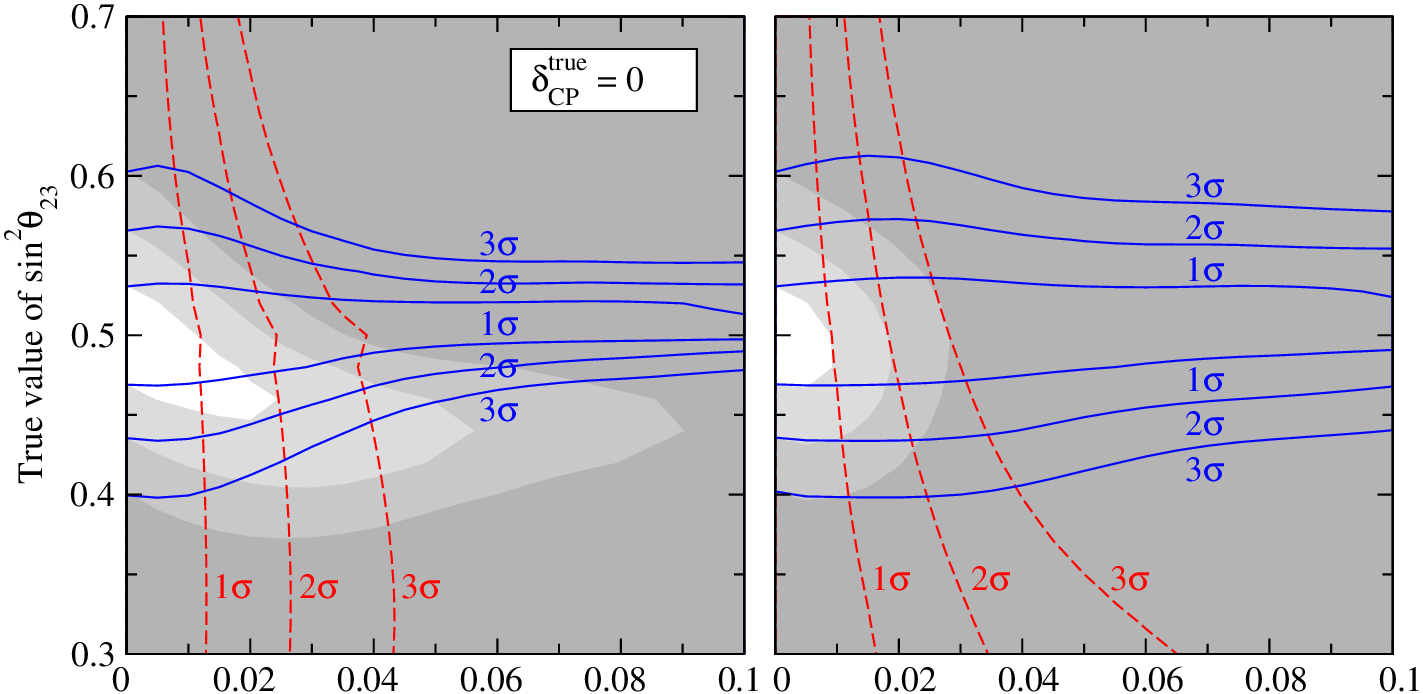}
   \includegraphics[width=0.58\textwidth]{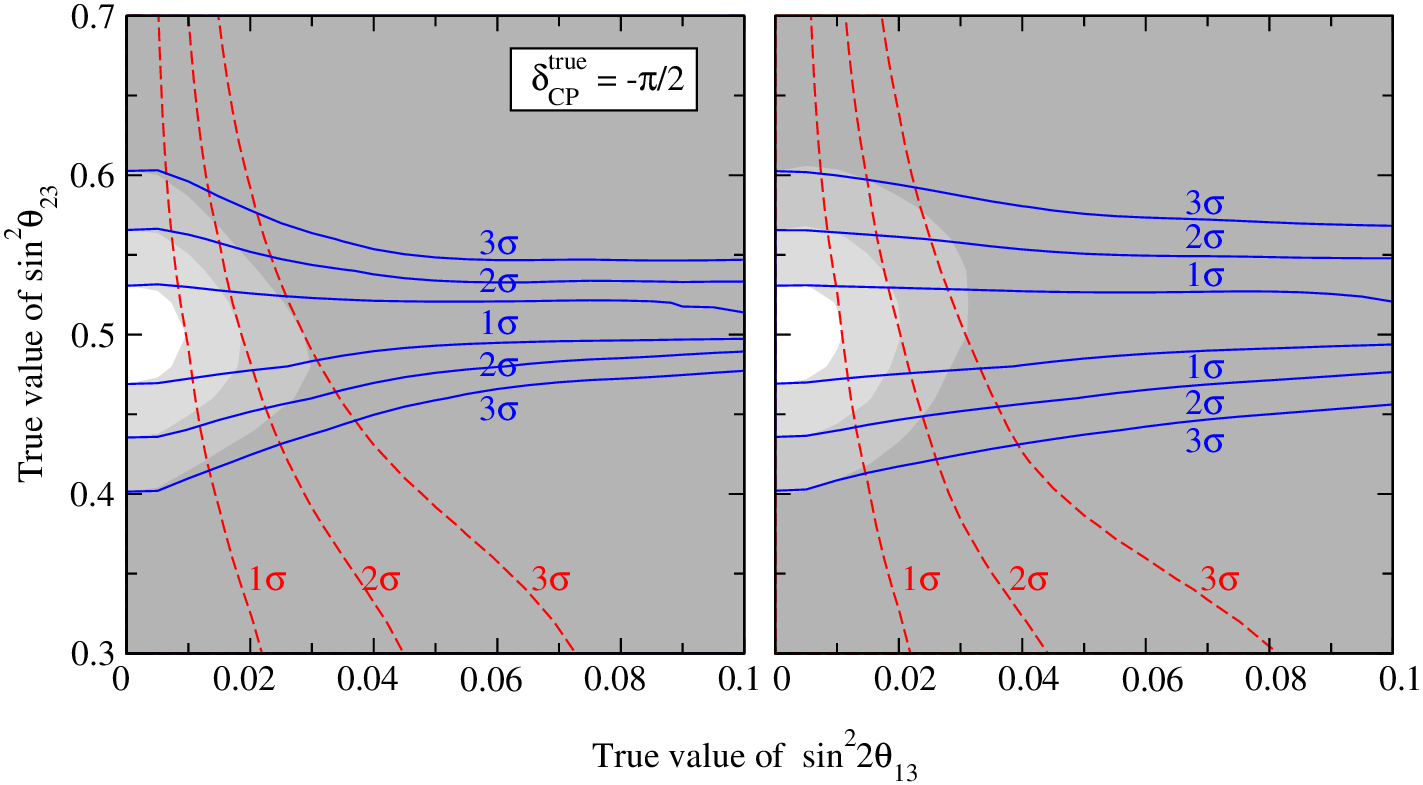}
   \mycaption{Rejection of the fake solutions from LBL+ATM data as a function
   of the true values of $\stheta$ and $\sin^2\theta_{23}$ for various
   $\deltacp^\mathrm{true}$ if the true hierarchy is normal (left) or inverted
   (right). Solid curves correspond to the solution with the wrong octant and
   the right hierarchy \owht, dashed curves to the right octant and the wrong
   hierarchy \othw, and shaded regions to the wrong octant and the wrong
   hierarchy \owhw. We show the contours of $\Delta\chi^2 = 1,4,9$ between the
   fake and the true solution, corresponding to a rejection of the fake
   solution at the $1\sigma$, $2\sigma$, and $3\sigma$ CL (from light to dark
   shading) for 1~dof. }
   \label{fig:resolve}
\end{figure}

The examples shown in \Fig~\ref{fig:d-th-nonmax-2} suggest that the LBL+ATM
combination offers an efficient method to reject the degenerate solutions.
To investigate this more systematically we have performed a scan over the true
values for $\theta_{13}$ and $\theta_{23}$ and various values for
$\deltacp^\mathrm{true}$. For a given point in the space of
$\stheta^\mathrm{true}$, $\sin^2\theta_{23}^\mathrm{true}$ and
$\deltacp^\mathrm{true}$ we test the ability to rule out each of the three
degenerate solutions \owht, \othw, and \owhw. This is done by minimizing
$\chi^2_\mathrm{LBL+ATM}$ with respect to all fit parameters, constraining the
octant of $\theta_{23}$ and the sign of $\dma$ corresponding to the fake
solution which we want to test. The results of this analysis are shown in
\Fig~\ref{fig:resolve}, which represents one of the main results of this work.

\begin{figure}[t!]
   \centering 
   \includegraphics[width=0.7\textwidth]{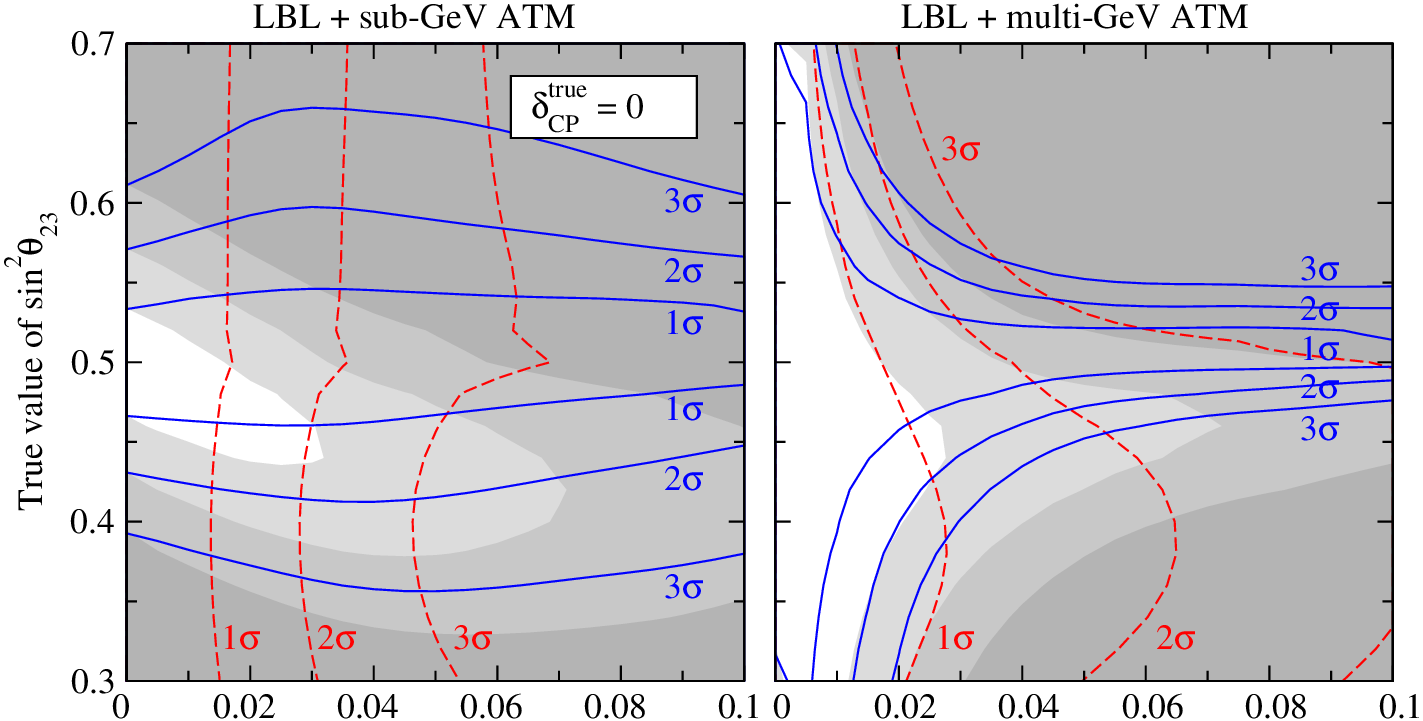}
   \includegraphics[width=0.7\textwidth]{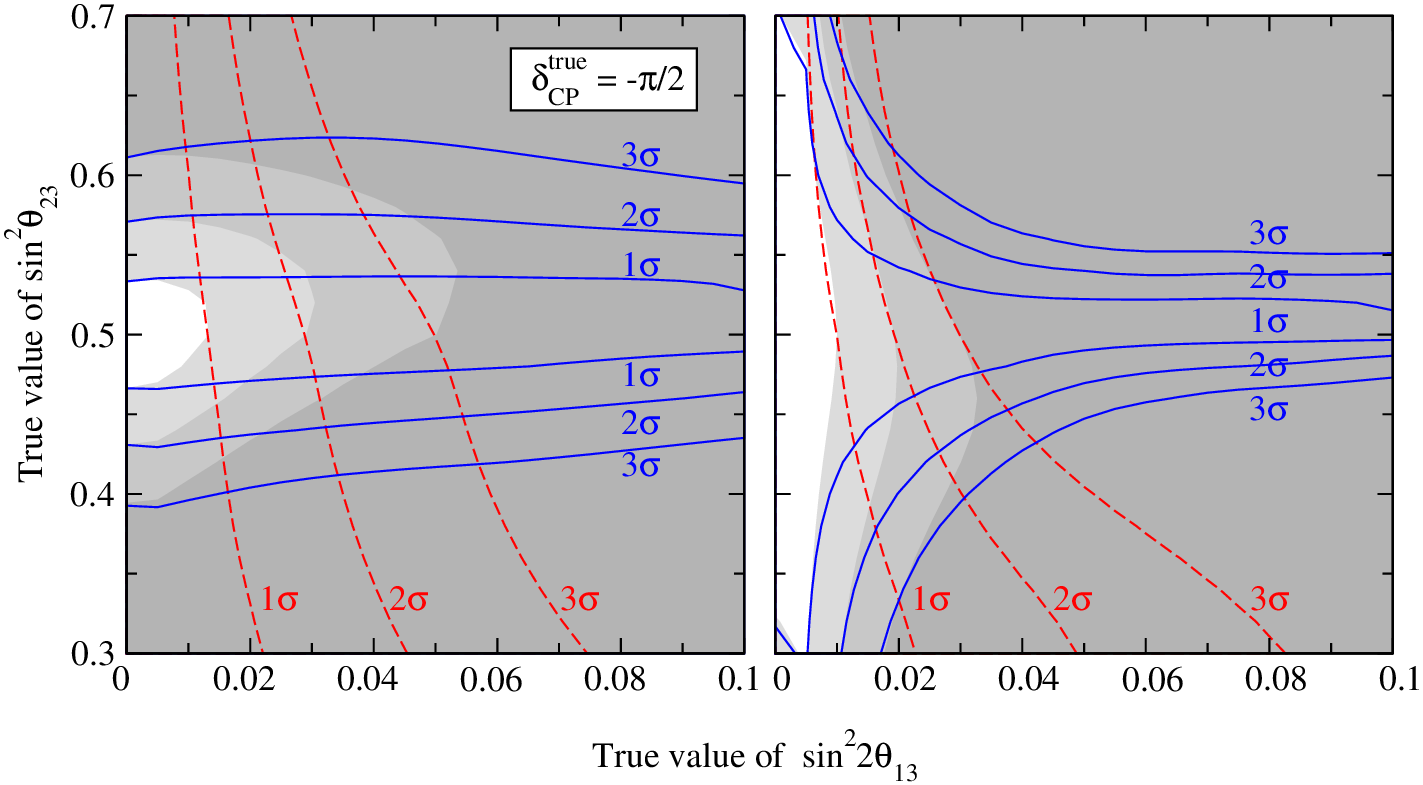}
   \mycaption{Rejection of the fake solutions from LBL data combined with
   sub-GeV (left) and multi-GeV (right) $e$-like and $\mu$-like ATM data as a
   function of the true values of $\stheta$ and $\sin^2\theta_{23}$ for
   $\deltacp^\mathrm{true} = 0$ (upper) and $\deltacp^\mathrm{true} =-\pi/2$
   (lower) and true normal hierarchy. Solid curves correspond to the solution
   with the wrong octant and the right hierarchy \owht, dashed curves to the
   right octant and the wrong hierarchy \othw, and shaded regions to the wrong
   octant and the wrong hierarchy \owhw. We show the contours of $\Delta\chi^2
   = 1,4,9$ between the fake and the true solution, corresponding to a
   rejection of the fake solution at the $1\sigma$, $2\sigma$, and $3\sigma$
   CL (from light to dark shading) for 1~dof.}
   \label{fig:resolve-atm-samples}
\end{figure}

The solid curves in \Fig~\ref{fig:resolve} show that the \owht-solution
corresponding to the $\theta_{23}$-degeneracy can be excluded at high
confidence level if $\theta_{23}$ is far enough from $\pi/4$. (Note that if
$\theta_{23}$ is close to $\pi/4$ this degeneracy disappears anyway.)  For
small values of $\stheta^\mathrm{true}$ this follows mainly from the
atmospheric sub-GeV $e$-like events, as discussed in
\Sec~\ref{sec:qual-atm-th23}. Moreover, if the true hierarchy is normal (left
panels of \Fig~\ref{fig:resolve}) we find an improvement of the octant
sensitivity for $\stheta \gtrsim 0.04$. This effect comes from the multi-GeV
$e$-like events, where resonant enhancement occurs for sufficiently large
$\stheta$, and therefore an additional dependence on $\sin^2\theta_{23}$
appears according to \eq~(\ref{eq:excess-multi}). This can be seen from
\Fig~\ref{fig:resolve-atm-samples}, where we show the ability to reject the
fake solutions using only sub-GeV (left) and multi-GeV (right) ATM data.
The improvement for large $\stheta$ is also visible for the inverted mass
ordering (right panels of \Fig~\ref{fig:resolve}), however, in that case the
effect is much smaller, since the resonance occurs for anti-neutrinos in the
inverted hierarchy, which contribute less to the total ATM data, and hence the
significance of the effect is smaller than in the normal hierarchy, where the
resonance occurs for neutrinos.

The wrong hierarchy solution \othw\ can be excluded to the right of the dashed
curves in \Fig~\ref{fig:resolve}. We observe that the ability to reject this
solution increases with $\sin^2\theta_{23}$. As discussed in
\Sec~\ref{sec:qual-atm-hier} this follows from the fact that the excess of
$e$-like multi-GeV events is enhanced for large $\sin^2\theta_{23}$. In
particular, for $\sin^2\theta_{23} > 0.5$ the sensitivity is completely
dominated by multi-GeV data (compare \Fig~\ref{fig:resolve-atm-samples}), and
we observe a very small dependence on the true value of $\deltacp$ and on the
true hierarchy.
For $\sin^2\theta_{23} < 0.5$ the situation becomes more complicated, and the
sensitivity depends on the true value of $\deltacp$. In that region also
sub-GeV data can contribute significantly, as visible in
\Fig~\ref{fig:resolve-atm-samples}. For low values of $\sin^2\theta_{23}$ the
effect of multi-GeV data becomes suppressed by the flux factor $(r
\sin^2\theta_{23} - 1)$ in \eq~(\ref{eq:excess-multi}), and the contribution
of sub-GeV data from \eq~(\ref{eq:excess-sub}) can become of comparable size.
Moreover, as discussed in \Ref~\cite{Peres:2003wd} for certain parameter
values an interference term between the two contributions may become relevant,
which depends on $\deltacp$. Therefore, the final sensitivity emerges from a
rather involved interplay of various effects and ATM data samples.

Finally, moving to the \owhw-solution, we find that in most cases this
``double'' degenerate solution can be excluded if either the \owht- or the
\othw-solution is ruled out. However, for certain parameter configurations an
interesting degeneracy between the octant and the mass ordering appears, such
that the \owhw-solution cannot be excluded, although the \owht- and the
\othw-solutions are not allowed. We find from \Fig~\ref{fig:resolve} that it
is difficult to distinguish a normal mass hierarchy and $\sin^2\theta_{23}
\simeq 0.45$, $\deltacp \simeq 0$ from an inverted hierarchy and
$\sin^2\theta_{23} \simeq 0.55$, $\deltacp \simeq \pi$. We observe from the
upper panels of \Fig~\ref{fig:resolve-atm-samples} that for these particular
parameters sub-GeV as well as multi-GeV data cannot exclude the
\owhw-solution. However, the appearance of this remaining degeneracy strongly
depends on the true value of $\deltacp$. For example, from the lower panels
of \Fig~\ref{fig:resolve-atm-samples} it becomes clear that it does not occur
for $\deltacp = -\pi/2$. The dependence on $\deltacp$ indicates that again the
interference term between $\dms$- and $\theta_{13}$-effects is important.


\section{Determining the neutrino mass hierarchy}
\label{sec:hierarchy}

In this section we discuss in detail the possibility to identify the type of
the neutrino mass ordering from a combined LBL+ATM analysis. To this aim we
simulate data for the \THK\ LBL experiment as well as for atmospheric neutrino
data assuming a ``true'' neutrino mass ordering. Then we fit these data with
the ``wrong'' hierarchy and search for the minimum $\chi^2$-value in the full
6-dimensional space of oscillation parameters, taking into account both
$\theta_{23}$-degenerate solutions \othw\ and \owhw. If this minimum is larger
than a certain value the wrong hierarchy can be excluded at the corresponding
CL.

\subsection{Sensitivity to the mass hierarchy for maximal $\theta_{23}$ mixing}
\label{sec:hier-max}

\begin{figure}[t!]
   \centering \includegraphics[width=0.95\textwidth]{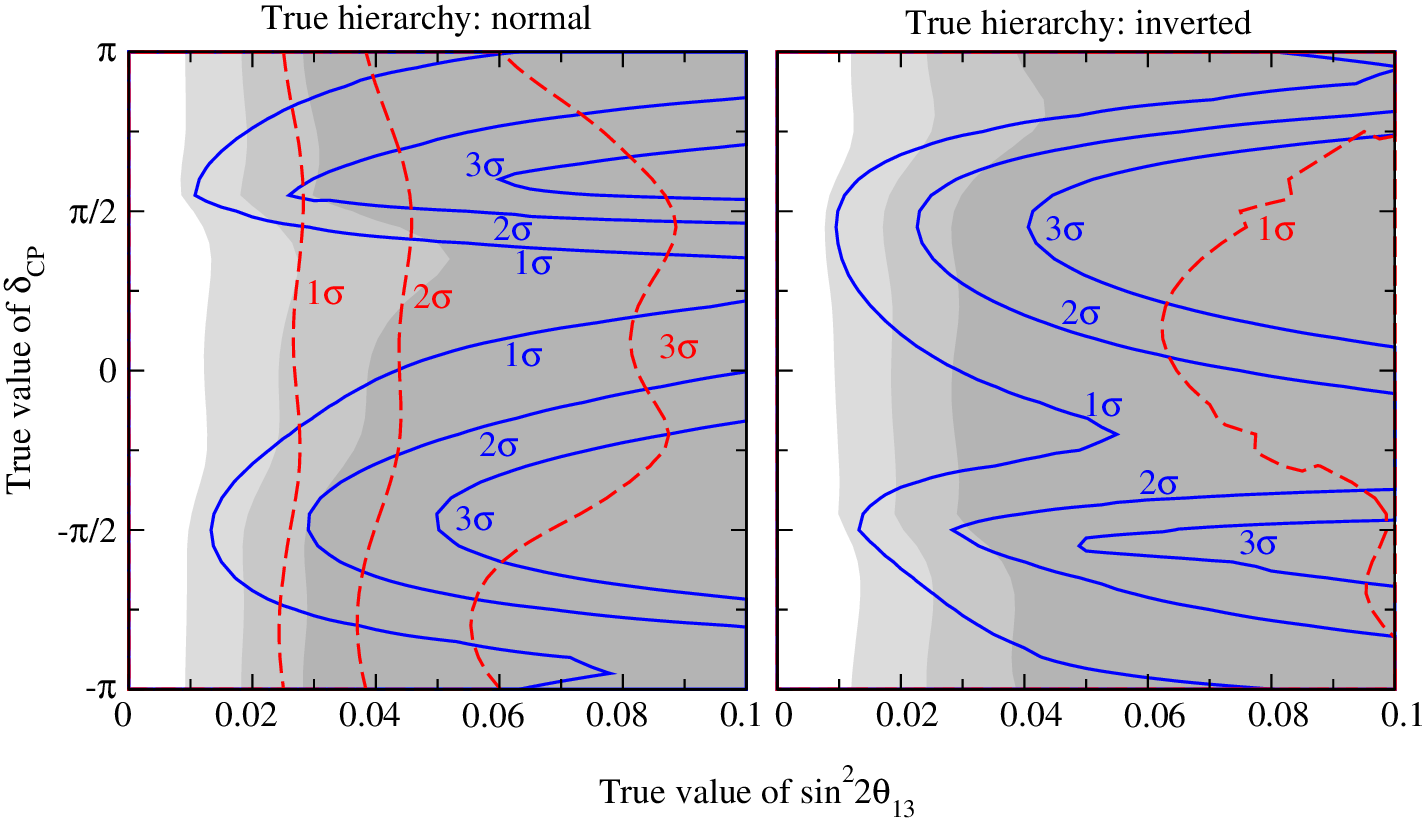}
   \mycaption{Sensitivity to the mass hierarchy as a function of the true
   values of $\stheta$ and $\deltacp$ for $\theta_{23}^\mathrm{true} = \pi/4$,
   if the true hierarchy is normal (left) or inverted (right). We show the
   contours of $\Delta\chi^2 = 1,4,9$ between the wrong and the true
   hierarchy, corresponding to a rejection of the wrong hierarchy at the
   $1\sigma$, $2\sigma$, and $3\sigma$ CL (from light to dark shading) for
   1~dof. The shaded regions correspond to LBL+ATM data combined, solid curves
   correspond to LBL-only, and dashed curves to ATM-only.}
   \label{fig:hierarchy}
\end{figure}

Before considering the general case we will first adopt the choice
$\theta_{23}^\mathrm{true} = \pi/4$, such that the $\theta_{23}$-degeneracy is
absent, and we are left only with the two-fold ambiguity related to
sign($\dma$). We show the sensitivity to the hierarchy in
\Fig~\ref{fig:hierarchy} as a function of the true values of $\stheta$ and
$\deltacp$. First we observe from this figure that the sensitivity of LBL data
alone strongly depends on the true value of $\deltacp$. For certain values of
$\deltacp$ the ability to distinguish normal and inverted hierarchy is even
lost for $\stheta = 0.1$. The main reason for the difficulties of LBL alone to
determine the mass hierarchy, is that for \THK\ the
matter effect is very small. ATM data on their own allow to identify the
normal mass hierarchy at $2\sigma$~CL for $\stheta^\mathrm{true} \gtrsim 0.04$
(see left panel of \Fig~\ref{fig:hierarchy}), whereas there is rather poor
sensitivity to the inverted hierarchy (right panel of
\Fig~\ref{fig:hierarchy}).  However, the sensitivity is significantly
increased by combining the two data sets: For LBL+ATM data the wrong hierarchy
can be excluded at $2\sigma$~CL for $\stheta^\mathrm{true} \gtrsim 0.02$, with
a rather small dependence on the true value of $\deltacp$ or the true type of
the hierarchy.

\begin{figure}[t!]
   \centering 
   \includegraphics[width=0.95\textwidth]{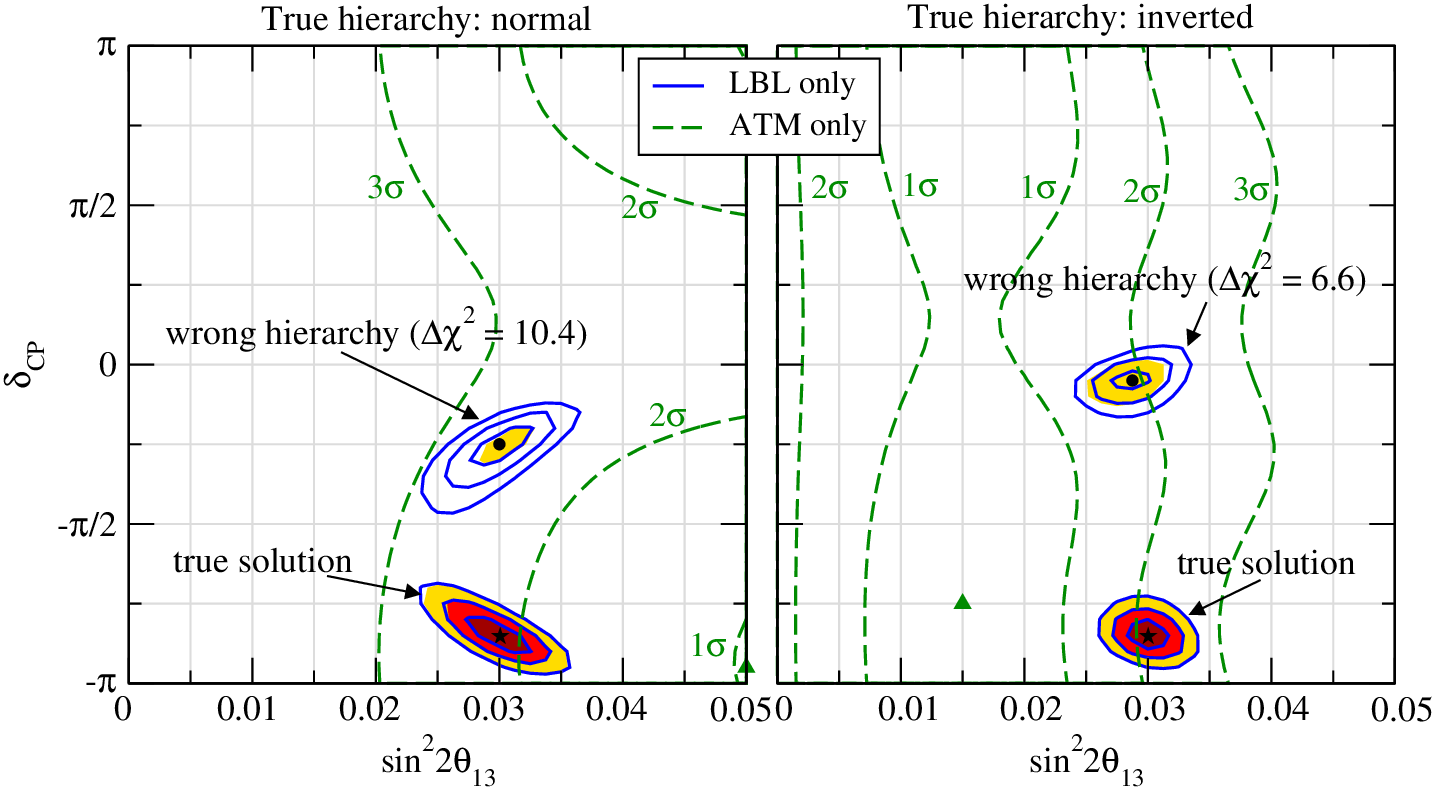}
   \mycaption{Allowed regions in the ($\stheta,\deltacp$) plane assuming for
   the true mass spectrum a normal (left) and an inverted (right) hierarchy,
   and $\theta_{23}^\mathrm{true} = \pi/4$. The true values
   $\stheta^\mathrm{true} = 0.03$ and $\deltacp^\mathrm{true} = -0.85\pi$ are
   marked by a star.  The shaded areas and the solid curves correspond to the
   1, 2, and 3 $\sigma$ allowed regions (2~dof) for the true and the
   sign($\Delta m^2_{31}$)-degenerate solution for LBL+ATM and LBL data,
   respectively. The dashed curves correspond to the $\Delta\chi^2$
   contours of ATM data for the wrong sign of $\Delta m^2_{31}$. The
   $\Delta\chi^2$ of the best fit point with the wrong hierarchy for LBL+ATM
   data is given in the figure and its location is marked by a dot. }
   \label{fig:delta-theta}
\end{figure}

These results can be understood qualitatively from the discussion given in
\Sec~\ref{sec:qual-atm-hier}.  To achieve the same value of
$\epsilon_e^\mathrm{multi}$ as implied by the (true) normal hierarchy and a
given $\stheta^\mathrm{true}$ with the inverted hierarchy, one has to adopt
values of $\stheta$ larger than $\stheta^\mathrm{true}$ to increase the effect
for the inverted hierarchy. Conversely, if the true hierarchy is inverted and
one wants to fit $\epsilon_e^\mathrm{multi}$ with the normal ordering,
$\stheta$ has to be smaller than $\stheta^\mathrm{true}$. This expectation is
confirmed in \Fig~\ref{fig:delta-theta}, where the allowed regions in the
($\stheta,\deltacp$) plane from ATM data are shown as dashed curves for the
true values $\stheta = 0.03$ and $\deltacp = -0.85\pi$. The left panel shows
that the normal hierarchy can be fitted with the inverted one for rather large
values of $\stheta$, whereas the opposite situation is visible for the
inverted hierarchy in the right panel.

This behavior explains also the ATM-only results shown in
\Fig~\ref{fig:hierarchy}. The reasonable sensitivity of ATM data to the normal
hierarchy visible in the left panel is given by the fact, that a fit with the
inverted hierarchy would be only possible for very large values of
$\stheta$. This is disfavored by data, first because the values of $\stheta$
required to fit the data start to get in conflict with the CHOOZ bound, and
hence are disfavored, and second because such large values of $\stheta$ start
to be excluded from ATM data on their own. On the other hand the reason for
the poor sensitivity to the inverted hierarchy shown in the right panel of
\Fig~\ref{fig:hierarchy} is that in general small values of
$\epsilon_e^\mathrm{multi}$ are expected, which can be easily fitted with the
normal hierarchy but smaller $\stheta$ (compare right panel of
\Fig~\ref{fig:delta-theta}), well below the sensitivity of ATM data.

Let us now discuss the effect of the sign($\dma$)-degeneracy for LBL data.  In
agreement with the discussion of \Sec~\ref{sec:qual-lbl} we observe from
\Fig~\ref{fig:delta-theta} that the sign($\Delta m^2_{31}$) degeneracy affects
mainly the determination of $\deltacp$, whereas the value of $\stheta$ of the
wrong-sign solution practically coincides with the true one.
According to \eq~(\ref{eq:fake-delta-h}) we expect for the fake
solution the value $\deltacp' = -0.15\pi$, which is in good agreement with the
values obtained in the actual fit: $\deltacp' = -0.25(-0.05)\pi$ for true
hierarchy normal (inverted).  The deviations from the value of $\deltacp'$
given by \eq~(\ref{eq:fake-delta-h}) are due to the matter effect,
which is not included in the probability \eq~(\ref{eq:emuprob}) and to some 
extent also to spectral information, which in general is difficult to include
in the analytical discussion.
Note that because of the choice $\theta_{23}^\mathrm{true} = \pi/4$ the \othw-
and \owhw-solutions coincide, and we are left only with the two-fold ambiguity
implied by the sign($\dma$)-degeneracy, as visible in
\Fig~\ref{fig:delta-theta}.

From \Fig~\ref{fig:delta-theta} also the reason for the significant
improvement of the sensitivity to the hierarchy for the LBL+ATM
combination becomes clear. Although for LBL data alone the degenerate
solution is present at a wrong value of $\deltacp$ even at
$1\sigma$~CL, the value of $\stheta$ has to be very similar to the
true one. This makes it impossible to accommodate ATM data with the
wrong hierarchy, since either values $\stheta$ significantly larger or
smaller than the true one are necessary to obtain the correct value
for $\epsilon_e^\mathrm{multi}$ with the wrong hierarchy. The location
in the ($\stheta,\deltacp$) plane required by the degenerate solution
of LBL gives for ATM data a $\chi^2$ corresponding to $2-3\sigma$~CL.
Therefore the combined data lead to the rather good sensitivity
implied by the $\Delta\chi^2$-values given in
\Fig~\ref{fig:delta-theta}. 

To summarize, the example shown in \Fig~\ref{fig:delta-theta} demonstrates how
the combination of LBL and ATM data leads to a sensitivity to the mass
hierarchy at the $3\sigma$ level, although each data set on its own cannot
distinguish normal from inverted mass ordering for this particular choice of
parameter values.  The sensitivity of ATM data alone suffers from the fact
that the wrong hierarchy can be accommodated by adjusting $\theta_{13}$, but
also the parameters $\theta_{23}$, $\dma$, and $\deltacp$ are important.  To
benefit from the earth matter effects in $e$-like events from atmospheric
neutrinos for the hierarchy determination the rather precise measurement of
the oscillation parameters from LBL data is mandatory. In particular, it is
necessary to know $\theta_{23}$ and $\dma$ at the 1\% level, and a reasonable
constraint on $\stheta$ is required.  
Finally, the determination of the correct mass hierarchy through the LBL+ATM
combination allows to resolve the ambiguity in $\deltacp$ from the
sign($\dma$)-degeneracy, and hence the potential to measure $\deltacp$ is
significantly increased. For the example shown in \Fig~\ref{fig:delta-theta}
the correct value of $\deltacp$ can be identified at $2\sigma$ CL.


\subsection{Sensitivity to the mass hierarchy for non-maximal
$\theta_{23}$ mixing}
\label{sec:hier-nonmax}

\begin{figure}[t!]
   \centering
   \includegraphics[width=0.95\textwidth]{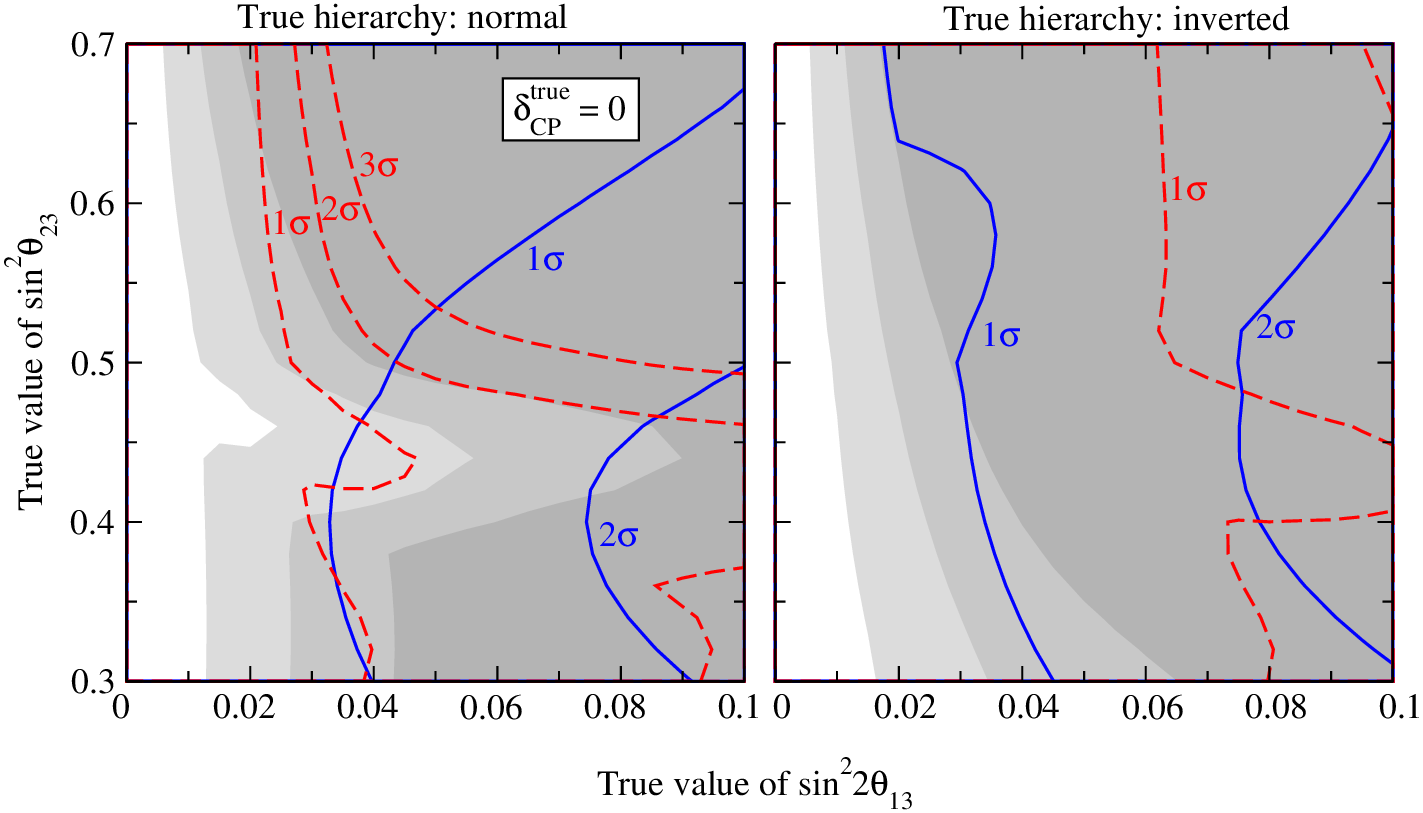}
   \mycaption{Sensitivity to the mass hierarchy as a function of the true
   values of $\stheta$ and $\sin^2\theta_{23}$ for $\deltacp^\mathrm{true} =
   0$ and a true normal (left panel) and inverted (right panel) hierarchy.  We
   show the contours of $\Delta\chi^2 = 1,4,9$ between the wrong and the true
   hierarchy, corresponding to a rejection of the wrong hierarchy at the
   $1\sigma$, $2\sigma$, and $3\sigma$ CL (from light to dark shading) for
   1~dof. The shaded regions correspond to LBL+ATM data combined, solid curves
   correspond to LBL-only, and dashed curves to ATM-only.}
   \label{fig:h.sq23}
\end{figure}

Now we generalize the discussion of the previous subsection to non-maximal
values of $\theta_{23}^\mathrm{true}$, such that in general all four solutions
of \eq~(\ref{eq:solutions}) are present. To identify the correct mass
hierarchy the \othw- as well as the \owhw-solution has to be excluded. In
\Fig~\ref{fig:h.sq23} we show the sensitivity to the mass hierarchy as a
function of the true values of $\stheta$ and $\sin^2\theta_{23}$ for LBL, ATM,
and LBL+ATM data. First we note that the results for LBL data-only strongly
depend on the true value of $\deltacp$, as observed already in
\Fig~\ref{fig:hierarchy}. ATM data on their own provide only a reasonable
sensitivity to the mass hierarchy if the true hierarchy is normal and
$\sin^2\theta_{23}^\mathrm{true} > 0.5$. This is the region of large excess of
$e$-like multi-GeV events (see discussion in \Sec~\ref{sec:qual-atm-hier}),
which cannot be achieved for any configuration in the inverted hierarchy.

The sensitivity to the hierarchy for the LBL+ATM combination can be inferred
from \Fig~\ref{fig:resolve} by considering the intersection of the shaded
region (\owhw-solution), with the region to the right of the dashed curves
(\othw-solution). The shaded regions in \Fig~\ref{fig:h.sq23} give an explicit
example for $\deltacp^\mathrm{true} = 0$. For the true inverted hierarchy
(right panel) the sensitivity to determine the mass hierarchy is given by the
ability to exclude the \othw-solution. As discussed in
\Sec~\ref{sec:degeneracies} the worsening of the sensitivity for low values of
$\sin^2\theta_{23}$ can be understood from the flux factor $(r
\sin^2\theta_{23} - 1)$ in the multi-GeV $e$-like event excess. If the true
hierarchy is normal (left panel) the \othw- as well as the \owhw-solutions
have to be taken into account. The ``spike'' visible in \Fig~\ref{fig:h.sq23}
at $\sin^2\theta_{23}^\mathrm{true} \sim 0.45$ comes from the fact that the
double degenerate solution \owhw\ cannot be excluded.

\begin{figure}[t!]
   \centering
   \includegraphics[width=0.6\textwidth]{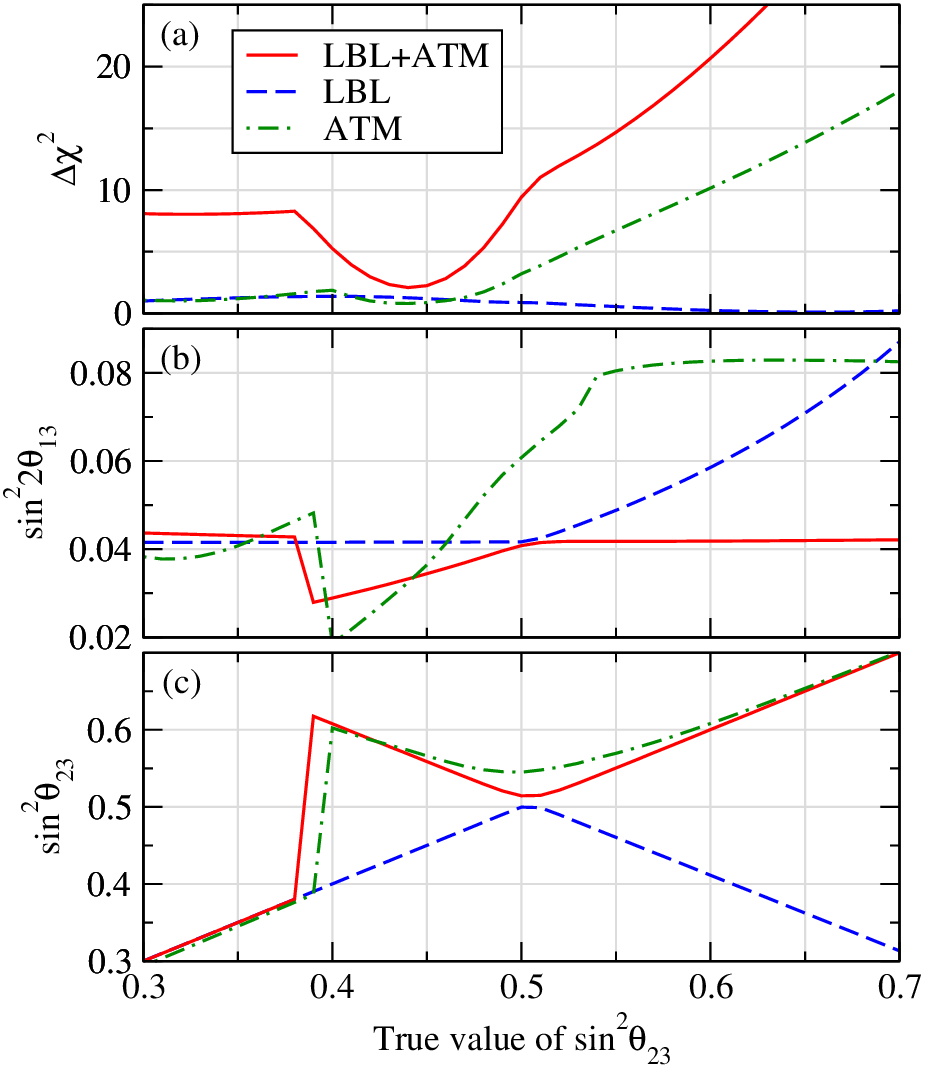}
   \mycaption{Panel~(a) shows the $\Delta\chi^2$ of the wrong hierarchy,
   panel~(b) and (c) show the values of $\stheta$ and $\sin^2\theta_{23}$ at
   the best fit point of the wrong hierarchy as a function of the true value
   of $\sin^2\theta_{23}$. The true hierarchy is normal and
   $\stheta^\mathrm{true} = 0.04$, $\deltacp^\mathrm{true} = 0$. }
   \label{fig:h-bfp}
\end{figure}

We illustrate this behavior in \Fig~\ref{fig:h-bfp}, where we show the
location of the best fit point for the wrong hierarchy as a function of the
true value of $\sin^2\theta_{23}$. We observe from panel~(c) that for
$\sin^2\theta_{23}^\mathrm{true} > 0.5$ the minimum for LBL+ATM follows the
true value of $\sin^2\theta_{23}$, \ie, it corresponds to the
\othw-solution. As discussed in \Sec~\ref{sec:degeneracies} this implies in
turn that $\stheta$ has to be close to the true value, as can be seen from
panel~(b). However, in that case the large values of
$\epsilon_e^\mathrm{multi}$ from the normal hierarchy cannot be obtained in
the inverted hierarchy, which leads to the large $\chi^2$ values, \ie\ good
sensitivity, in that region.
For $0.4 < \sin^2\theta_{23}^\mathrm{true} < 0.5$ one observes from
\Fig~\ref{fig:h-bfp} that the best fit moves to the \owhw-solution,
characterized by the wrong $\theta_{23}$, see panel~(c), and a value of
$\stheta$ given by \eq~(\ref{eq:fake-th13-th23}), see panel~(b). As mentioned
in \Sec~\ref{sec:resolve} the occurrence of this degeneracy strongly depends
on the true value of $\deltacp$, and follows from a delicate interplay of
effects in sub- and multi-GeV data.
In the region of $\sin^2\theta_{23}^\mathrm{true} < 0.4$ the best fit returns
again to the \othw-solution, characterized by the true $\theta_{23}$ and a
value of $\stheta$ close to the true one. From \Fig~\ref{fig:h-bfp} we find
that the fit of LBL-only, as well as ATM-only chooses best fit values of
$\theta_{23}$ and $\stheta$ close to the true ones. This suggests that in that
region the sensitivity comes from effects related to $\dma,\dms$, and
$\deltacp$. As mentioned in \Sec~\ref{sec:resolve} here the main sensitivity
comes from the ATM sub-GeV data sample (compare
\Fig~\ref{fig:resolve-atm-samples}) and the interference term between $\dms$-
and $\dma$-effects~\cite{Peres:2003wd} can become important, which introduces
the dependence on $\deltacp$.

\begin{figure}[t!]
   \centering
   \includegraphics[width=0.95\textwidth]{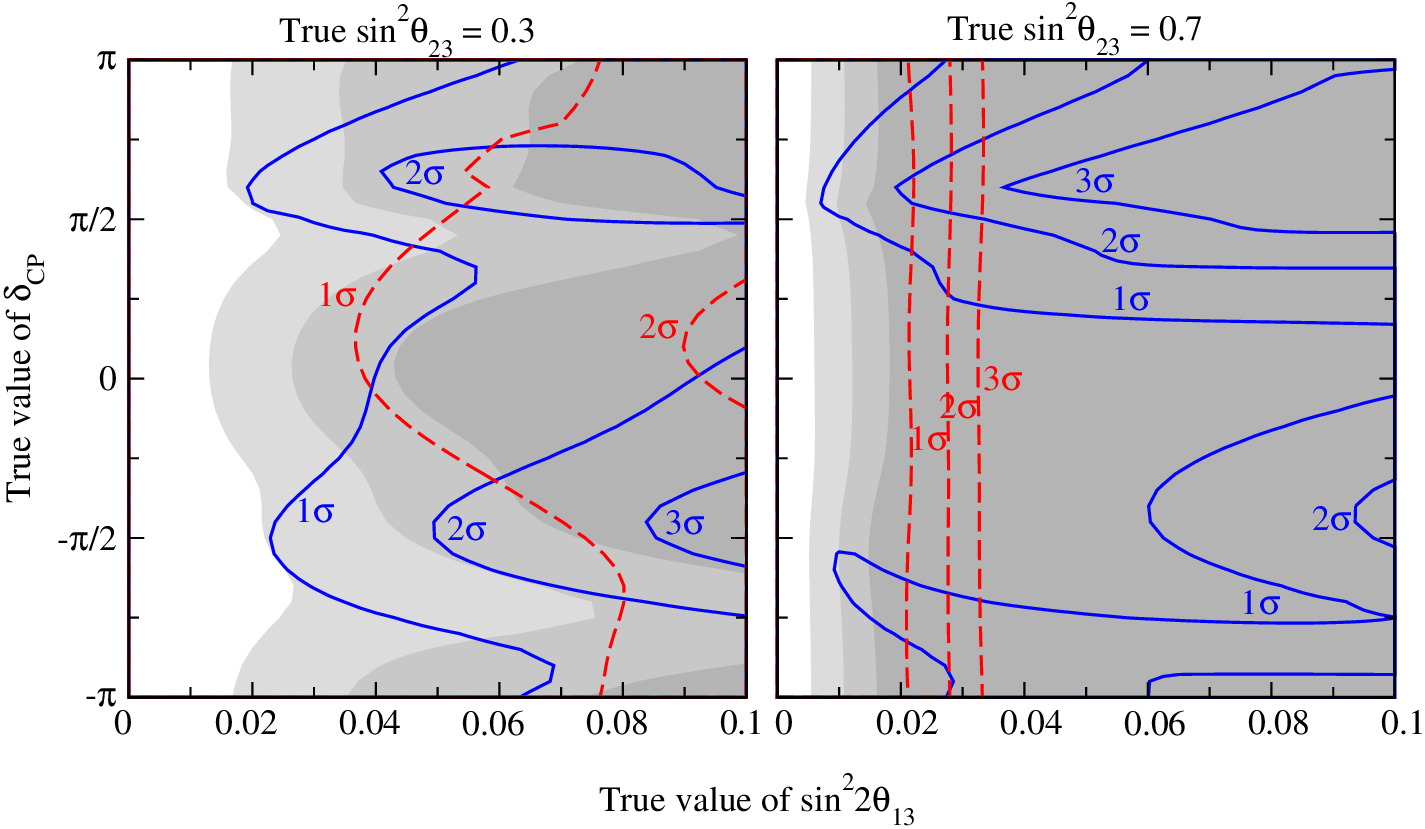}
   \mycaption{Sensitivity to the mass hierarchy as a function of the true
   values of $\stheta$ and $\deltacp$ for $\sin^2\theta_{23}^\mathrm{true} =
   0.3$ (left) and $\sin^2\theta_{23}^\mathrm{true} = 0.7$ (right). The true
   hierarchy is normal.  We show the contours of $\Delta\chi^2 = 1,4,9$
   between the wrong and the true hierarchy, corresponding to a rejection of
   the wrong hierarchy at the $1\sigma$, $2\sigma$, and $3\sigma$ CL (from
   light to dark shading). The shaded regions correspond to LBL+ATM data
   combined, solid curves correspond to LBL-only, and dashed curves to
   ATM-only.}
   \label{fig:h-nonmax}
\end{figure}

The strong dependence of the sensitivity on $\deltacp$ for small
$\sin^2\theta_{23}$ is visible in the left panel of \Fig~\ref{fig:h-nonmax},
where we show the sensitivity to the hierarchy as a function of
$\stheta^\mathrm{true}$ and $\deltacp^\mathrm{true}$ for
$\sin^2\theta_{23}^\mathrm{true} = 0.3$. The rich structure visible for the
LBL+ATM combination indicates a complicated interplay and/or cancellations of
various effects.
On the other hand, from the right panel we observe that for large values of
$\sin^2\theta_{23}^\mathrm{true}$ the sensitivity to the hierarchy becomes
practically independent of the true value of $\deltacp$. In this range
multi-GeV $\dma$-effects are enhanced by the flux factor and dominate over
$\dms$-effects and the interference term. Also ATM data alone provide quite a
good sensitivity, thanks to the big effect in multi-GeV $e$-like events.


\section{The $\theta_{23}$-degeneracy and the 
sensitivity to $\sin^22\theta_{13}$}
\label{sec:octant-th13}

As discussed in \Sec~\ref{sec:degeneracies}, the presence of the octant
degeneracy leads to severe complications in the determination of $\stheta$.
In this section we consider the specific case $\theta_{13}^\mathrm{true} =
0$, and we investigate how the upper bound on $\stheta$ (``sensitivity to
$\stheta$'') can be improved by the combined LBL+ATM analysis. The fact that
the $\theta_{23}$-degeneracy affects the $\stheta$-sensitivity is well-known,
see \eg, \Refs~\cite{Barger:2001yr,Huber:2002mx,Donini:2003vz,Yasuda:2004gu}.
The choice $\theta_{13}^\mathrm{true} = 0$ essentially eliminates the
differences between normal and inverted hierarchies, \ie\ the degeneracy is
perfect, and the fact that it cannot be resolved introduces only very small
ambiguities for other parameters. Hence the \owht- and \owhw-solutions
coincide within good accuracy. Moreover, the analysis is independent of the
true value of $\deltacp$, since the phase becomes unphysical for $\theta_{13}
= 0$.  We define the sensitivity to $\stheta$ as the largest value of
$\stheta$ which fits the data generated for a true value
$\stheta^\mathrm{true}=0$ at a given CL (see Appendix~C of
\Ref~\cite{Huber:2004ug} for a detailed discussion).

\begin{figure}[t!]
   \centering
   \includegraphics[width=0.95\textwidth]{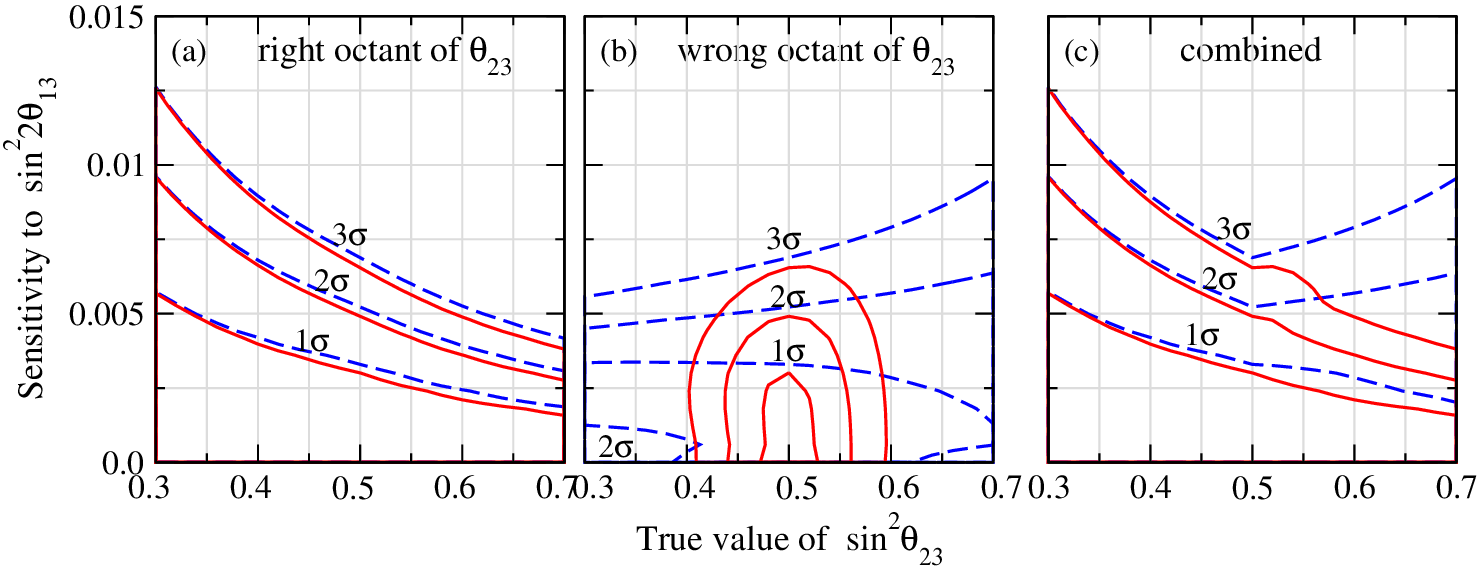}
   \mycaption{Sensitivity to $\stheta$ as a function of the true value
   of $\sin^2\theta_{23}$ for LBL data only (dashed), and the
   combination LBL+ATM (solid). In panels (a) and (b) we restrict the
   fit of $\theta_{23}$ to the octant corresponding to
   $\theta_{23}^\mathrm{true}$ and $\pi/2 -
   \theta_{23}^\mathrm{true}$, respectively, whereas panel (c) shows
   the overall sensitivity taking into account both octants.}
   \label{fig:th13sens}
\end{figure}

Let us first assume that the true octant of $\theta_{23}$ was known.  In that
case one expects from \eq~(\ref{eq:emuprob}) that the sensitivity for
$\stheta$ becomes better for increasing $\sin^2\theta_{23}$, since for large
$\sin^2\theta_{23}$ the first term in \eq~(\ref{eq:emuprob}), which is
proportional to $\stheta$, is enhanced. Moreover, the $\alpha^2$ term gets
suppressed for small $\cos^2\theta_{23}$, which reduces the effect of
multi-parameter correlations induced by $\dms$. This expectation is confirmed
in panel~(a) of \Fig~\ref{fig:th13sens}, where the sensitivity to $\stheta$ is
shown as a function of the true value of $\sin^2\theta_{23}$, assuming that
the octant is known.
However, if the octant of $\theta_{23}$ is not known, the degenerate
solution prevents the smaller $\stheta$-limits implied by
$\sin^2\theta_{23}^\mathrm{true} > 0.5$. In that case the data can be
fitted by relatively large values of $\stheta$, since exchanging
$\sin^2\theta_{23}$ and $\cos^2\theta_{23}$ reduces the effect of
$\stheta$ and increases the $\alpha^2$ term in \eq~(\ref{eq:emuprob}).
This effect is shown by the curves corresponding to LBL data in
panel~(c) of \Fig~\ref{fig:th13sens}. We observe that the sensitivity
to $\stheta$ gets worse for non-maximal values of
$\theta_{23}^\mathrm{true}$ in both octants because of the presence of
the degenerate solution.

Since the octant-degeneracy can be efficiently resolved by atmospheric data
one expects a significant improvement of the $\stheta$-sensitivity for the
combined LBL+ATM analysis. In panel~(b) of \Fig~\ref{fig:th13sens} we show the
$\stheta$-limit constraining the fit to the {\it wrong} octant of
$\theta_{23}$. As expected from \Fig~\ref{fig:wrong-th23} there are no
solutions for LBL+ATM data if $\theta_{23}$ is sufficiently far from maximal,
since the wrong solution is disfavored by ATM data. Consequently, we find
that the final result shown in panel~(c) for LBL+ATM is very close to the
situation in panel~(a), where only the true solution for $\theta_{23}$ has
been used.
Hence, the $\stheta$-sensitivity is significantly increased for $\theta_{23} >
\pi/4$ through the exclusion of the wrong octant solution by ATM data. There
is no relevant improvement for $\theta_{23} < \pi/4$, since in that case the
sensitivity to $\stheta$ suffers for the true $\theta_{23}$ solution because
of the $\stheta$-suppression implied by the small values of
$\sin^2\theta_{23}$.


\section{Conclusions}
\label{sec:conclusions}

In this work we have performed a combined analysis of future long-baseline
(LBL) and atmospheric (ATM) neutrino data. As a specific example we have
considered the phase~II of the T2K experiment, consisting of a 4~MW superbeam
produced at the J-PARC facility. The 1~Mt HyperKamiokande water Cherenkov
detector will serve as detector for the LBL experiment and simultaneously
provide high statistics ATM neutrino data.
We have shown that the combined LBL+ATM analysis offers a very appealing
possibility to resolve parameter degeneracies in the LBL data. In particular,
the ambiguities implied by the sign($\dma$)- and the
$\theta_{23}$-degeneracies can be lifted to a large extent. This becomes
possible through three-flavor effects in ATM data related to $\theta_{13}$ and
$\dms$. A systematic scan of the parameter space has been performed to
investigate the ability to determine the type of the neutrino mass
ordering. Let us summarize our main findings:
\begin{itemize}
\item
For true values of $\sin^2\theta_{23} > 0.5$ the correct mass hierarchy can be
identified at $2\sigma$~CL if $\stheta \gtrsim 0.015$, rather independent of
the true value of $\deltacp$ or the type of the true hierarchy. In this region
the sensitivity is dominated by multi-GeV $e$-like events in ATM data.
\item
For true values of $\sin^2\theta_{23} < 0.5$ the correct mass hierarchy can be
identified at $2\sigma$~CL if $\stheta \gtrsim 0.03$, where here the actual
sensitivity depends on the true value of $\deltacp$ and the true
hierarchy. The final sensitivity emerges from an interplay of various
effects in the different ATM data samples.
\item
The solution with the wrong octant of $\theta_{23}$ can be excluded at
$3\sigma$~CL if $|\sin^2\theta_{23} - 0.5| \gtrsim 0.1$, independent of the
true values of $\theta_{13}$, $\deltacp$, and the hierarchy.  This follows
mainly from $\dms$-effects in $e$-like sub-GeV ATM data. If $\stheta \gtrsim
0.03$ and the true hierarchy is normal the octant sensitivity at $3\sigma$~CL
improves to $|\sin^2\theta_{23} - 0.5| \gtrsim 0.05$ due to
$\theta_{13}$-effects in the multi-GeV ATM data.
\item
The lifting of the degeneracies by ATM data significantly increases the
performance of the LBL experiment for the measurement of $\stheta$ and
$\deltacp$, since fake solutions implied by the degeneracies can be ruled out.
Generically, the determination of the correct octant of $\theta_{23}$ removes
an ambiguity in the measurement of $\stheta$, whereas lifting the
sign($\dma$)-degeneracy allows the identification of the correct value of
$\deltacp$.
\item
If no finite value for $\theta_{13}$ is found the upper limit on $\stheta$ is
significantly improved by resolving the $\theta_{23}$-degeneracy by ATM data
if $\theta_{23} > \pi/4$.
\end{itemize}
Let us stress that these results follow from the complementarity of the two
data sets; neither LBL data alone nor ATM data alone can provide a comparable
physics reach.
The LBL data allow a very precise determination of $|\dma|$ and
$\sin^22\theta_{23}$, and although $\theta_{13}$ and $\deltacp$ suffer from
ambiguities related to the degeneracies they are constrained to rather
specific values. However, there is very poor sensitivity to the mass hierarchy
and to the octant of $\theta_{23}$ for LBL data alone.
For ATM data alone only for $\sin^2\theta_{23}^\mathrm{true} \gtrsim 0.6$ and
a true normal hierarchy a reasonable sensitivity exists for the mass ordering,
since in this case a big excess of $e$-like multi-GeV events is predicted,
which cannot be achieved by any configuration within the inverted
hierarchy. For all other regions in the parameter space the sensitivity of
ATM-only is rather poor, since the data can be fitted with the wrong hierarchy
by adjusting the oscillation parameters.
To benefit from the hierarchy sensitivity offered by $\theta_{13}$-earth
matter effects in ATM data the precise measurement of the oscillation
parameters from LBL data is mandatory. In particular, it is necessary to know
$\theta_{23}$ and $|\dma|$ at the 1\% level, and a reasonable constraint on
$\stheta$ is required. Hence, only the {\it combination} of LBL and ATM data
makes it possible to obtain a good sensitivity to the neutrino mass hierarchy.
Let us add that to benefit from the LBL+ATM combination indeed a few
$\mathrm{Mt\,yrs}$ of ATM data are necessary. In particular we have checked
that for exposures below $1\,\mathrm{Mt\,yrs}$ the sensitivity to the hierarchy
is essentially lost.

Finally, let us remark that the methods to resolve parameter degeneracies
discussed previously in general involve the combination of two or more
(expensive) experiments, \eg\ at different baselines or using different
oscillation channels. In contrast, once a LBL experiment with a megaton
Cherenkov detector is built, ATM data come for free. Therefore, in addition to
the interesting physics, the synergies between LBL and ATM data offer a
relatively economical way of resolving parameter degeneracies.

\subsection*{Acknowledgment}

M.M.\ is supported by the National Science Foundation grant PHY-0354776.
T.S.\ is supported by the VI~Framework Programme of the European Community
under a Marie Curie Intra-European Fellowship.


\end{document}